\documentclass[aps,prx,reprint,amsmath,amssymb,floatfix,longbibliography,superscriptaddress]{revtex4-1}

\usepackage{graphicx}
\usepackage{amsmath,amssymb,bbold,bm,color}
\usepackage{float}
\usepackage{epstopdf}
\usepackage{hyperref}
\usepackage[dvipsnames]{xcolor}
\usepackage[normalem]{ulem}

\newcommand{\bB}{{\bm B}}

\newcommand{\cT}{{\cal T}}
\newcommand{\cK}{{\cal K}}

\newcommand{\cG}{{\cal G}}
\newcommand{\cF}{{\cal F}}

\newcommand{\cO}{{\cal O}}

\newcommand{\bee}{\begin{equation}}
\newcommand{\ee}{\end{equation}}

\newcommand{\ket}[1]{| #1 \rangle}

\hypersetup{colorlinks=true,linkcolor=blue,citecolor=blue,urlcolor=blue}

\begin{document}

\title{Diagnosing quantum chaos in many-body systems using entanglement as a resource}

\author{\'{E}tienne Lantagne-Hurtubise}
\email{lantagne@phas.ubc.ca}
\affiliation{Department of Physics and Astronomy \& Quantum Matter Institute, University of British Columbia, Vancouver BC, Canada V6T 1Z4}
\affiliation{Kavli Institute for Theoretical Physics, University of California Santa Barbara, CA 93106, USA}
\author{Stephan Plugge}
\email{plugge@phas.ubc.ca}
\affiliation{Department of Physics and Astronomy \& Quantum Matter Institute, University of British Columbia, Vancouver BC, Canada V6T 1Z4}
\author{Oguzhan Can}
\affiliation{Department of Physics and Astronomy \& Quantum Matter Institute, University of British Columbia, Vancouver BC, Canada V6T 1Z4}
\author{Marcel Franz}
\affiliation{Department of Physics and Astronomy \& Quantum Matter Institute, University of British Columbia, Vancouver BC, Canada V6T 1Z4}
\affiliation{Kavli Institute for Theoretical Physics, University of California Santa Barbara, CA 93106, USA}

\date{\today}

%
\begin{abstract} 
Classical chaotic systems exhibit exponentially diverging trajectories
due to small differences in their initial state. The analogous
diagnostic in quantum many-body systems is an exponential growth of
out-of-time-ordered correlation functions (OTOCs). These quantities can be
computed for various models, but their experimental study
requires the ability to evolve quantum states backward in time, similar to
the canonical Loschmidt echo measurement. In some simple systems, backward time evolution can be achieved by reversing the sign of the Hamiltonian; however in most interacting many-body systems, this is not a viable option. 
Here we propose a new family of protocols for OTOC measurement that do not require backward time evolution. Instead, they rely on ordinary time-ordered
measurements performed in the thermofield double (TFD)
state, an entangled state formed between two identical copies of the
system. We show that, remarkably, in this situation the Lyapunov chaos exponent $\lambda_L$ can be extracted from the measurement of an ordinary two-point correlation function.
As an unexpected bonus, we find that our proposed method yields the so-called ``regularized'' OTOC -- a  quantity that is believed to most directly indicate quantum chaos. According to recent theoretical work, the TFD state can be prepared as the ground state of two weakly coupled identical systems and is therefore amenable to  experimental study.  
We illustrate the utility of these protocols on the example of the maximally chaotic Sachdev-Ye-Kitaev model and support our findings by extensive numerical simulations.   
\end{abstract}

\date{\today}
\maketitle

\section{Introduction}
A key characteristic of chaotic quantum many-body systems is rapid
dispersal of quantum information deposited among a small
number of elementary degrees of freedom. After a short time the information is distributed 
among exponentially many degrees of freedom, whereby it becomes effectively lost to all local observables.
This apparent loss of quantum information through unitary evolution, known also as ``scrambling'', lies at the heart of thermalization in closed systems and plays a key role in understanding quantum aspects of black holes as epitomized by Hawking's information loss paradox \cite{Hawking1976}.
Black holes are believed to scramble at the fastest possible rate consistent with
causality and unitarity \cite{Hayden2007}. Some strongly coupled quantum systems, such
as the Sachdev-Ye-Kitaev (SYK) model \cite{SY1996,Kitaev2015,Maldacena2016,Sachdev2015} and its variants, are also known to be fast scramblers; this motivates their description as duals of gravitational theories containing a black hole. 
Scrambling in quantum theories can be quantified through the out-of-time order correlators defined below, which -- for chaotic systems -- show exponential
growth at intermediate times with a characteristic Lyapunov exponent $\lambda_L$.
A universal upper bound on chaos, conjectured by Maldacena, Shenker and Stanford \cite{Maldacena2016b}, posits that $\lambda_L\leq 2\pi T$ and is saturated in the class of maximally chaotic systems which includes black holes and SYK models.

Diagnosing quantum chaos and scrambling in realistic physical systems
is a problem of fundamental importance that has been only partially
addressed so far. As we review below, the great hurdle in conventional
approaches to diagnosing chaos is the necessity to evolve the quantum
system backward in time during the measurement \cite{Swingle2016,Grover2016,Swingle2018}.
So far this has been achieved in a very limited range of systems, mainly quantum simulators \cite{Du2017} and ion traps \cite{Rey2017,Monroe2019}.
However there is no hope of applying this method to a broader class of naturally occurring quantum many-body systems, because in those one simply does not possess the level of control required to reverse the time evolution.
In this paper we address this pressing challenge by introducing a new approach to diagnosing chaos in quantum many-body systems which does not require backward time evolution during the measurement.
The approach is based on 
a procedure which creates an entangled resource state that permits
chaos diagnosis through an {\em ordinary} measurement.
 Specifically, we show that in this resource state the chaos exponent $\lambda_L$ governs
  the exponential growth of an ordinary two-point correlator at
  intermediate times and can thus be experimentally accessed
using  routine spectroscopic techniques.

Rather than a Loschmidt echo, our scheme more closely resembles the approach of Refs.~\cite{Daley2012,Abanin2012,Islam2015} to the measurement of Renyi entropy by readout of entangling operators between identical copies of a quantum state. Alternative approaches for the detection of scrambling in quantum systems include interferometry~\cite{Yao2016arxiv} or out-of-equilibrium measurement protocols~\cite{Campisi2017,Halpern2017}.
In our case, the entanglement between identical copies of the chaotic quantum system is generated as resource from a specifically engineered Hamiltonian. The ability to detect OTOCs hence emerges from measurements of \emph{simple} operators in an a-priori \emph{complicated} ground state of two coupled chaotic systems.

A quantitative measure of scrambling in quantum systems is given by the
expectation value of a commutator squared \cite{Shenker2014}, 
\begin{equation}\label{e1}
C_\eta(t)=-\langle [W(t),V(0)]_\eta^2\rangle
\end{equation}
of two initially commuting Hermitian operators $[W,V]_\eta=0$. In the
following we allow for both bosonic ($\eta=1$) and fermionic
($\eta=-1$) statistics, with $[\: \cdot \:, \: \cdot \:]_\eta$ denoting
the commutator (anti-commutator) for $\eta=1$ ($\eta=-1$). The operators
evolve in time according to the system Hamiltonian $H$ through
$W(t)=e^{iHt}We^{-iHt}$ and  $\langle ...\rangle$ denotes the thermal
average at inverse temperature $\beta=1/T$. The intuition
behind the definition \eqref{e1} is the following: As the operator
$W(t)$ evolves in time it becomes more and more complex until it
eventually fails to commute with operator $V$. One thus expects $C_\eta(t)$
to grow as a function of time and eventually saturate at a value close
to $2\langle V^2\rangle \langle W^2\rangle$ for large $t$, regardless
of the specific form of $V$ and $W$. In chaotic many-body systems, at
intermediate times, the growth of $C_\eta(t)$ follows an exponential
dependence, as long as $V$ and $W$ are ``simple'' operators composed of products of a small number of elementary degrees
of freedom.

Expanding the commutator in Eq.\ \eqref{e1} we obtain two types of
thermal averages
\begin{eqnarray}\label{e2}
C_\eta(t)&=& \eta \langle W(t)VVW(t)\rangle+ \eta \langle VW(t)W(t)V\rangle \\
&-& \langle VW(t)VW(t)\rangle-\langle W(t)V W(t)V\rangle. \nonumber
\end{eqnarray}
The averages on the first line represent naturally time-ordered
correlators (NTOC) that correspond to ``sensible'' experiments
performed in an ordinary quantum system. For instance
the second term  $\langle VW(t)W(t)V\rangle$ describes a process in
which we perturb the system at time $t=0$ by applying operator $V$,
evolve the perturbed system forward in time and then perform a
measurement of a quantity represented by operator $W^2$.

The averages on the second line  of Eq.\ \eqref{e2} represent
out-ot-time-ordered correlators. These correspond to less sensible
experiments. The second term for instance, which is often denoted as
\begin{equation}\label{e3}
F(t)=\langle W(t)V W(t)V\rangle,
\end{equation}
describes the process in which we compare two states of the system:
one obtained by first perturbing with $V$, then applying $W(t)$ at a later time $t$; the other obtained by perturbing first with $W(t)$,
evolving {\em backward} in time, then applying $V$. We note that in the special case where operators $V$ and $W$ are also unitary, the quantity $C_\eta(t)$ can be expressed
simply as 
\begin{equation}\label{e4}
C_\eta(t)= 2 \eta -2 {\rm Re}F(t).
\end{equation}

\begin{figure}[t]
	\includegraphics[width = 8.5cm]{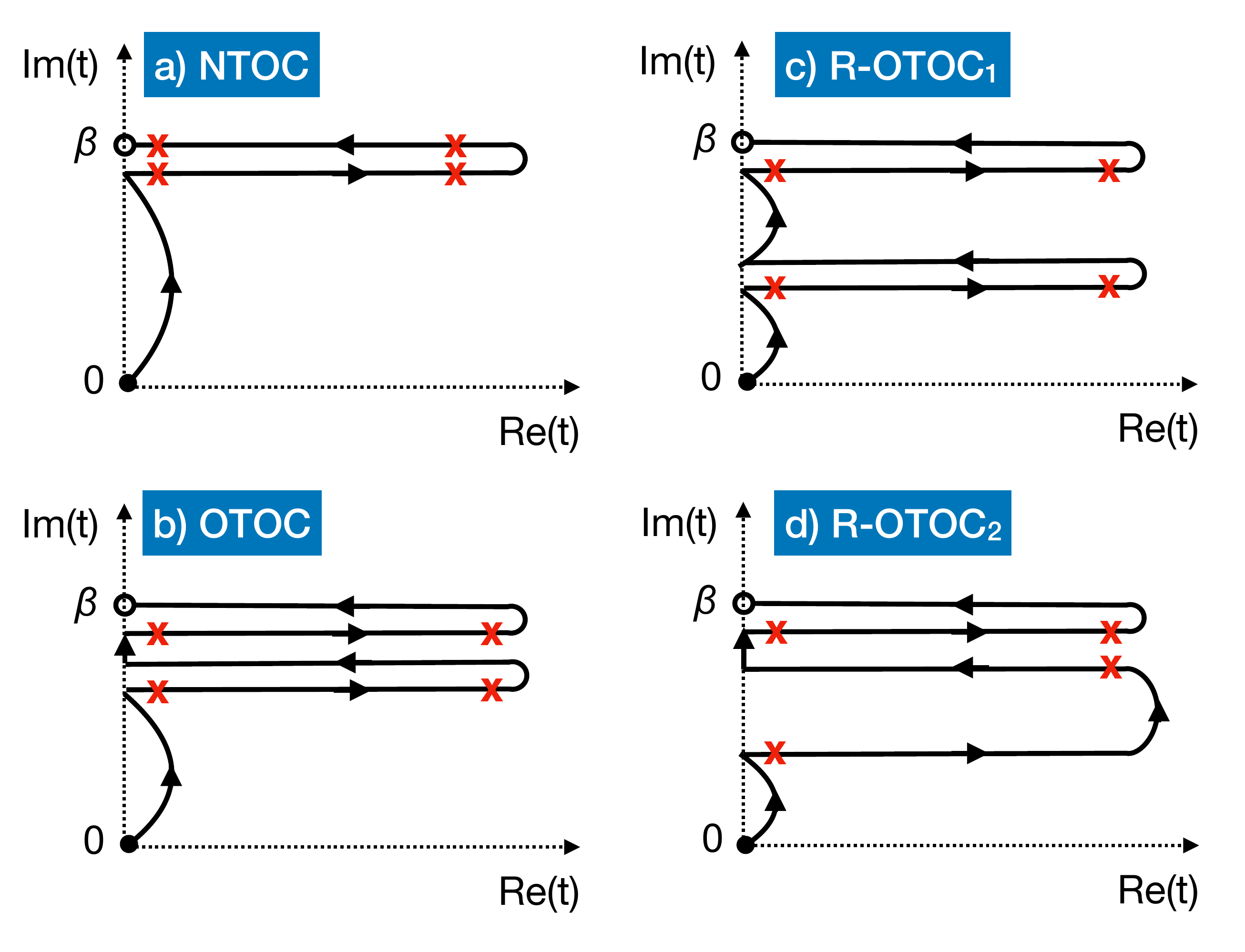}
	\caption{Naturally time-ordered vs.\ out-of-time-ordered correlation
		functions on the Schwinger-Keldysh contour. Evolution in real time
		$t$ follows horizontal lines, whereas imaginary time evolution
		(generated by powers of the thermal density matrix $e^{-\beta H}$)
		occurs along the vertical direction. Positions of the operators $W$
		and $V$ are marked by crosses. Panels (a) and (b) show NTOC and
		OTOC, respectively, with conventional placement of the density
		matrix. Panels (c) and (d) show two different ``regularized'' OTOCs
		discussed in Section II.
	}\label{fig1}
\end{figure}
The fundamental  difference between NTOC and OTOC is best visualized
by placing the operators on the Schwinger-Keldysh contour
illustrated in Fig.\ \ref{fig1}. The imaginary time evolution is
generated by the density matrix which becomes explicit if we rewrite
Eq.\ \eqref{e3} as $F(t)={\rm tr}[W(t)V W(t)V e^{-\beta H}]$. Notice
that while NTOC can be represented by placing the operators on a
conventional  Schwinger-Keldysh contour with one forward and one
backward evolving branch  (Fig.\ \ref{fig1}a), OTOC require a doubled
contour indicated in Fig.\ \ref{fig1}b.

As a rule, the physical implementation of an OTOC measurement
requires (actual or effective) backward time evolution and is therefore difficult to achieve
in most systems. Similar to Loschmidt echo
\cite{Hahn1950,Peres1984,Jalabert2001,Gorin2006},  a  measurement of
$F(t)$ is possible in situations where one controls the Hamiltonian at
the microscopic level and can, in particular, reverse its sign to
generate backward time evolution. As a practical matter this
restriction greatly limits the types of systems in which the phenomenon of
scrambling can be experimentally probed.

In the rest of the paper we introduce, discuss, and put to the test a family of
protocols that probe OTOCs but do not require backward time
evolution. They instead require two copies of the system prepared in a
special entangled state called ``thermofield double'' (TFD). As we explain
in detail below, TFD is a pure quantum state whose reduced density matrix
coincides with the thermal density matrix of one copy of the
system. The TFD state has been widely studied in quantum gravity theories as
a description of traversable wormholes \cite{Gao2017,Yang2017,Qi2018,Gao2019}.
It has a remarkable property, which we review below, that time effectively flows in
the opposite direction in two entangled systems. It is this property which underlies its usefulness in the proposed OTOC measurement protocols. Importantly, recent theoretical work has established a simple method that can be used to prepare the TFD
state \cite{Cottrell2019}.
The method consists of weakly coupling two identical subsystems in a specific way, then cooling the combined system to its ground state. As an example, the ground state wavefunction of two identical SYK Hamiltonians coupled by simple bilinear tunneling terms has $>96\%$ overlap with a TFD wavefunction \cite{Qi2018} (see also Fig.~\ref{fig2}a).
We also note interesting recent works on how to prepare a TFD state using quantum circuits~\cite{Hsieh2018,Martyn2018,Zhu2019}; however, these approaches are limited to the moderate system sizes accessible on present-day quantum simulators, similar to the experiments of Refs.~\cite{Du2017,Rey2017,Monroe2019}.

In the following we first review the concept of the TFD state, and
then demonstrate how it can be used to diagnose quantum chaotic
behavior via an OTOC measurement that does not require explicit
backward time evolution of quantum states (Sec.~\ref{sec:OTOC_TFD})
We discuss in detail how the TFD state can be prepared and used to
extract the chaos exponent $\lambda_L$ from an equilibrium measurement
of a two-point correlation function.  We then apply these general ideas to a pair of coupled SYK Hamiltonians, recently argued to be holographically dual to a \emph{traversable wormhole}, and known to admit a TFD ground state (Sec.~\ref{sec:SYK}).
This simple model serves as a testbed for demonstrating the usefulness of our OTOC measurement protocols. Finally, in Sec.~\ref{sec:physical_realizations} we discuss possible physical realizations of the coupled SYK models in the laboratory, and expand on the challenges of performing the necessary measurements.
We conclude with an outlook onto interesting future work and outstanding challenges in Sec.~\ref{sec:outlook}.

\section{OTOC measurement using the thermofield double state}
\label{sec:OTOC_TFD}

In this Section we review the concept of  the thermofield double state,
discuss its properties, and then show how it can be used to measure
OTOCs in a way that does not require explicit backward time evolution.

\subsection{TFD state: Definition and properties}
Consider two copies of the same system, left and right, described by
many-body Hamiltonians $H_L$ and $H_R$, respectively. We assume that $H_\alpha$ ($\alpha=L,R$) are invariant
under time reversal generated by an antiunitary operator $\Theta$. The
TFD state at inverse temperature $\beta$ is then defined as
\begin{equation}\label{h1}
|{\rm TFD}_\beta\rangle ={1\over \sqrt{Z_\beta}}\sum_n e^{-\beta E_n/2}|\bar{n}\rangle_L\otimes|n \rangle_R
\end{equation}
where $|n \rangle_\alpha$ is an eigenstate of $H_\alpha$ with
energy eigenvalue $E_n$, and $Z_\beta=\sum_n e^{-\beta E_n}$ is the partition function.
$|\bar{n}\rangle=\Theta|n\rangle$ denotes the time-reversed partner of
the eigenstate $|n\rangle$  which shares the same energy
eigenvalue $E_n$. We note that time reversal is necessary here to
define a unique TFD state, as each eigenstate $|n\rangle$ is defined up to an overall phase $e^{i\phi_n}$. A direct product
$|\bar{n}\rangle_L\otimes|n \rangle_R$ is however well-defined because $|\bar{n}\rangle$ transforms with the opposite
phase to $|n \rangle$ under the corresponding $U(1)$  transformation.
In the limit of zero temperature $|{\rm TFD}_\beta\rangle$
simply becomes a direct product of $L$ and $R$ ground states, whereas
at infinite temperature it becomes a maximally entangled state between
the two subsystems.

The TFD state has several important properties. The expectation value of
any one-sided operator with respect to $|{\rm TFD}_\beta\rangle$  
is given by a thermal average,
\begin{equation}\label{h1a}
\langle \cO_L\rangle_{\rm TFD} =Z_\beta^{-1}\sum_n e^{-\beta
	E_n} ~_L\langle n|\cO_L|n\rangle_L. 
\end{equation}
It is also important to note that  
$|{\rm TFD}_\beta\rangle$ is {\em not} an eigenstate of the full system
Hamiltonian $H=H_L+H_R$. It is, however, an eigenstate with eigenvalue
zero of $H_-=H_L-H_R$;  it can be easily checked that 
\begin{equation}\label{h2}
(H_L-H_R)|{\rm TFD}_\beta\rangle=0.
\end{equation}
This has implications for the concept of time-translation invariance
in the TFD state. Eq.\ \eqref{h2} implies that $|{\rm
	TFD}_\beta\rangle$ evolves trivially under $H_-$, 
\begin{equation}\label{h2a}
e^{-it(H_L-H_R)} |{\rm
	TFD}_\beta\rangle =|{\rm TFD}_\beta\rangle.
\end{equation}
Hence the expectation value of a product of two operators acting in the $L$ and $R$ systems has the property
\begin{eqnarray}\label{h3}
\cF(t_1,t_2)&=&\langle \cO_L(t_1) \cO_R(t_2)\rangle_{\rm TFD} \nonumber \\
&=& \langle \cO_L(t_1+t) \cO_R(t_2-t)\rangle_{\rm TFD},
\end{eqnarray}
valid for arbitrary $t$. The second line follows upon replacing $|{\rm
  TFD}_\beta\rangle$ in the expectation value on the first line by
$e^{-it(H_L-H_R)} |{\rm  TFD}_\beta\rangle$ using Eq.~\eqref{h2a}, and recalling that $\cO_R$ commutes with $H_L$ (and same for $L\leftrightarrow R$).
Choosing $t=t_2$ we see that $\cF(t_1,t_2)$ is a function of $t_1+t_2$ only. 
This should be compared to the statement of time-translation
invariance in a conventional (unentangled) state, $\langle \cO(t_1)
\cO(t_2)\rangle=\langle \cO(t_1-t) \cO(t_2-t)\rangle$, where the expectation
value only depends on $t_1-t_2$ as long as the Hamiltonian is
independent of time. 

 In the context of wormhole physics, Eq.~\eqref{h3}
can be interpreted as time flowing {\em in the opposite direction} on
the two sides of the wormhole, represented in the quantum theory by two
entangled subsystems. It is this peculiar property that
ultimately allows one to use a TFD as a resource for OTOC measurement
without explicit backward time evolution.

\subsection{Probing OTOC using TFD state}

For the purposes of this subsection, we will assume that
we have the ability to engineer two identical copies of an interesting
quantum 
many-body system described by Hamiltonians $H_L$ and $H_R$ and
prepare them in the TFD state Eq.~\eqref{h1}. In the subsequent
Sections, we will discuss how this can be achieved in
practice, and study some concrete examples.
Here we focus on elucidating how a two-sided measurement performed on the TFD state can be used to probe correlation functions that map onto thermal OTOCs with respect to one subsystem.

Consider a naturally time-ordered correlator 
\begin{equation}\label{h4}
\tilde{F}(t,t')=\langle \cT [V_L(t)W_R(t) V_R(t')W_L(t')]\rangle_{\rm TFD} 
\end{equation}
evaluated with respect to the TFD state in Eq.~\eqref{h1}. Here $\cT$ denotes the
time-ordering operator. Normal time-ordered expectation values of
this type correspond to physical quantities that are measurable, at
least in principle. The specific average defined in Eq.~\eqref{h4} can be
thought of as a component of the current-current correlator (where the
current operator involves both sides of the composite system) that
would arise in the calculation of the appropriate linear response
conductance.  

We now assume $t>t'$ and write the average in Eq.~\eqref{h4}
explicitly using the TFD state~\eqref{h1}. We obtain 
\begin{multline}\label{h5}
\tilde{F}(t,t')={1\over Z_\beta}\sum_{n,m}e^{-\beta(E_n+E_m)/2} 
~_L\langle \bar{n}|V_L(t)W_L(t')|\bar{m}\rangle_L \\
\times ~_R\langle n|W_R(t)V_R(t')|m\rangle_R,
\end{multline}
where we used the fact that $L$ and $R$ operators (anti-)commute at all times,
and that they act only on $L$ and $R$ eigenstates, respectively. The right hand side has
been written as a product of expectation values taken in $L$ and $R$
systems separately. Because each such expectation value is a complex
number and the two subsystems are identical, we can now drop the $L$ and $R$
subscripts, recognizing that it does not matter in
which subsystem they are evaluated
\footnote{For fermionic systems, there is an additional subtlety in identifying the L and R operators. In order to preserve the fermionic commutation relations, one needs to define (say) $V_L \equiv V \otimes \mathcal{P}_R$ and $V_R \equiv I_L \otimes V$ (similarly for $W_L$ and $W_R$), where $\mathcal{P}_R$ is the fermionic parity operator acting on the $R$ system. Whenever a two-sided correlator contains an even number of fermionic operators on each side (such as in Eq. (10)), the $\mathcal{P}_R$ factors cancel out because $\mathcal{P}_R^2= I$. In more general cases one has to keep track of this factor, e.g. in deriving the analog of Eqs. (17) and (23) for fermionic operators.} 
We thus have 
\begin{multline}\label{h6}
\tilde{F}(t,t')={1\over Z_\beta}\sum_{n,m}e^{-\beta(E_n+E_m)/2} 
\langle m|W(-t')V(-t) |n\rangle \\
\times \langle n|W(t)V(t')|m\rangle,
\end{multline}
where we used general properties of time-reversed states \footnote{Here we assume that time-reversal takes the form $\Theta = \mathcal{K}$ where $\mathcal{K}$ is complex conjugation. In general time-reversal can also include a unitary part $\mathcal{U}$, $\Theta = \mathcal{U} \mathcal{K}$, in which case our discussion still applies with the appropriate insertions of the unitary in the time-ordered correlators.}, namely $\langle\bar{
	a}|\bar{b}\rangle=\langle\Theta
a|\Theta b\rangle=\langle b|a \rangle$, and
\begin{equation}\label{h6b}
\langle\bar{n}| \cO(t)|\bar{m}\rangle=
\langle m|\cO(-t)^\dagger|n\rangle.
\end{equation}
As the final step we insert the Boltzmann factors into the expectation values, and replace them by powers of the density matrix,  e.g. $ e^{-\beta(E_n+E_m)/2} \langle
n|W(t)V(t')|m\rangle=Z_\beta\langle n|y^2W(t)V(t')y^2|m\rangle$, with 
\begin{equation}\label{h7}
y^4=e^{-\beta H}/Z_\beta~.
\end{equation}
This allows us to perform the sum over $n$ using the completness
relation $\sum_n|n\rangle\langle n| =1$ and arrive at  the result
\begin{equation}\label{h8}
\tilde{F}(t,t')={\rm tr}[ W(-t')V(-t)y^2 W(t)V(t')y^2].
\end{equation}
Note that the trace is now evaluated with respect to the eigenstates
of a {\em single-sided} Hamiltonian. 

For any $t>0$ and $t'<0$ Eq.\ \eqref{h8} has the structure of an OTOC. In the
special case $t'=-t$ it becomes
\begin{equation}\label{h9}
\tilde{F}(t,-t)= {\rm tr}[ W(2t)V(0)y^2 W(2t)V(0)y^2],
\end{equation}
where we used the time-translation invariance to shift all
temporal arguments by $+t$ for clarity.

We observe that  $\tilde{F}(t,-t)$ coincides with the canonical OTOC function
$F(2t)$ defined in Eq.\ \eqref{e3}, except for the placement of
the density matrix powers $y$ (see also Fig.\ \ref{fig1}c).  
Expressions of this type are called ``regularized'' OTOCs and have been extensively studied in the literature \cite{Maldacena2016,Maldacena2016b}.
Regularized OTOCs exhibit less singular behavior than unregularized OTOCs when evaluated analytically, and they have been argued to more reliably measure quantum chaos in many-body systems \cite{Galitsky2018,Romero2019,Kobrin2020}. The universal upper bound on the Lyapunov exponent $\lambda_L$ has also been proven only for regularized OTOCs \cite{Maldacena2016b}. We see that, remarkably, naturally time-ordered correlators evaluated in the TFD state map onto {\em regularized} OTOCs with respect to the single system.

Following the sequence of steps between Eqs.\ \eqref{h4} and \eqref{h8}
it is possible to derive other useful identities that relate
expectation values of operators in the TFD state to single-sided OTOCs. For instance we
find
\begin{multline}\label{h10}
\langle W_L(-t) V_L(0)V_R(0)W_L(-t)\rangle_{\rm TFD} =\\
={\rm tr}[ W(t)V(0) W(t)y^2V(0)y^2]. 
\end{multline}
The first line can be interpreted as creating an excitation in the TFD
state at time $-t$, evolving forward in time and performing a
two-sided measurement of a quantity represented by the operator  $V_LV_R$ at time zero. This correlator also maps onto the
canonical OTOC albeit with a different regularization, illustrated in
Fig.\ \ref{fig1}d. Below we refer to this as ``asymmetric''
regularized OTOC.
We note that similar relations have been anticipated in the high-energy community \cite{Shenker2014, Shenker2015, Maldacena2016b}.

\subsection{Initial state preparation}
Our considerations above establish formal identities relating
naturally time-ordered correlators
evaluated in an entangled state of two identical systems 
to out-of-time-ordered correlators in the single system, such as Eqs.~\eqref{h9} and
\eqref{h10}. A sensible measurement performed in the TFD state can
thus provide information on the OTOC and diagnose
quantum chaotic behavior in a many-body system.  We now
discuss a method that allows one to prepare the TFD state. In
addition, we see from the discussion in the previous subsection that in
order to probe the OTOC one in fact needs $|{\rm
  TFD}_\beta\rangle$ at a {\em negative} time, for only then is the
correlator on the left hand side of Eq.\ \eqref{h10}  naturally
time ordered. The same remark applies to $\tilde{F}(t,t')$ defined in
Eq.\ \eqref{h4}. In the following we therefore discuss how a state closely
approximating $|{\rm  TFD}_\beta(-t)\rangle$ can be prepared in a
realistic setup.

The easiest way to prepare a TFD state in the laboratory would be to engineer a Hamiltonian $H_S$ which admits $|{\rm  TFD}_\beta \rangle$ as a \emph{ground state}. A collection of such Hamiltonians were recently constructed in Ref. \cite{Cottrell2019}, with a unique  $|{\rm  TFD}_\beta \rangle$ ground state separated from the rest of the spectrum by a gap of order $1/\beta$.
Thus, a TFD state can be prepared by engineering the system to obey Hamiltonian $H_S$, and then cooling it down to a physical temperature $T_\text{phys}$ small compared to the gap.
The form of $H_S$ required to obtain the TFD ground state {\em exactly} is complicated and therefore not practical from the standpoint of generating the state in a laboratory. However, the ground state $|\Psi_0\rangle$ of a simple Hamiltonian 
\begin{equation}\label{h12}
H_S=H_L+H_R+H_I,
\end{equation}
with 
%
%
\begin{equation}\label{h13}
H_I=\sum_j c_j d_j^\dagger d_j ~ , ~ d_j = \cO_L^j- \Theta \left(\cO_R^j\right)^\dagger \Theta^{-1}
\end{equation}
and $\Theta$ representing the time-reversal operator, has been shown \cite{Cottrell2019}  
to approximate $|{\rm  TFD}_\beta\rangle$ to good
accuracy for appropriately chosen coefficients $c_j$. Here
$\cO^{L/R}_j$ are arbitrary (but identical) operators acting on the $L$ and
$R$ systems, respectively. This method is expected to apply to generic many-body Hamiltonians $H_L$ respecting the eigenstate thermalization hypothesis \cite{Cottrell2019}, and is thus of broad relevance in the study of quantum chaotic systems.

In another recent work, Maldacena and Qi \cite{Qi2018} used an even simpler construction with a coupling of the form
\begin{equation}\label{eq:coupling_Qi}
H_I = i\mu \sum_j \cO_L^j \cO_R^j
\end{equation}
to describe an eternal traversable wormhole formed by two copies of the SYK model.
We will review this construction in the next Section, verify numerically that it admits a ground state that closely approximates the TFD state, and discuss some of its intriguing properties.

The above procedure allows one to prepare a state that closely
approximates $|{\rm  TFD}_\beta\rangle$ by coupling two systems through
$H_I$ defined in Eqs.~\eqref{h13} or \eqref{eq:coupling_Qi}, then cooling the combined system to reach its ground state. When the coupling $H_I$ is switched off
the system begins to evolve forward in time according to the decoupled
Hamiltonian 
\begin{equation}\label{h12a}
H_0=H_L+H_R.
\end{equation}
This evolution is non-trivial because $|{\rm  TFD}_\beta\rangle$ is not an eigenstate of $H_0$. In order to probe OTOC, however, we require a TFD state prepared at {\em negative} time $-t$, as discussed above.
It would thus appear that our protocol requires backward time evolution after all. We show in Appendix \ref{App:negativetimes}, that the required resource state $|\Psi_0(-t)\rangle\simeq |{\rm TFD}_\beta(-t)\rangle$ for short time durations $t$ can be prepared by manipulating the strength of the coupling $H_I$,
without the need to reverse the sign of $H_0$.
Since the ability to introduce and control  $H_I$ is necessary to
prepare the TFD state in the first place, this method does not
introduce any substantial additional complications.

\subsection{OTOC from two-point functions}

Here we discuss an  approach that allows one to extract the OTOC from the measurement of a time-ordered \emph{two-point} function $G_{LR}(t,t')$, in a generic chaotic system with coupling $\mu$ between the two sides given by Eq.~\eqref{eq:coupling_Qi}.
This method relies only on the fact that the
ground state of the coupled system closely approximates the TFD state and,
importantly, does not require varying $\mu$ before or during the measurement. 
In contrast to related previous work
\cite{Gharibyan2019, Vermersch2019}, our method comprises the measurement of a
single (averaged) Green's function, rather than statistics on an
ensemble of measurements. We outline the argument below and
provide technical details in Appendix \ref{AppA}.

We consider the two-point time-ordered $LR$ correlation function in real time,
\begin{equation}\label{s1}
iG_{LR}(t,t')=\langle
\cT V_L(t) V_R(t')\rangle.
\end{equation}
The average is taken with respect to the (TFD) ground state of the coupled system and is therefore time-translation invariant, $G_{LR}(t,t')=G_{LR}(t-t')$.
The operators inside the average evolve according to the full coupled Hamiltonian $H_S=H_L+H_R + H_I$.

At weak coupling  $\mu$ (compared to the energy scale $J$ of the chaotic Hamiltonian $H_0$), and for short time durations  $\mu |t-t'|\ll 1$, it is possible to rewrite the two-point correlator in Eq.~\eqref{s1} as a single-sided thermal average of operators evolving according to $H_L$ (or, equivalently, $H_R$).
Formally, this is done by passing from the Heisenberg picture to the interaction picture and expanding the corresponding time-evolution operator $U(t,t')$ to leading order in the small parameter $\mu |t-t'|$.
Details of this calculation are given in Appendix~\ref{AppA}, and the result is
\begin{eqnarray}
iG_{LR}(t,-t)&\simeq& {\rm tr}[Vy^2Vy^2] \label{ss2} \\
&-& 2\eta\mu\sum_j\int_0^tds \ {\rm
	tr}[\cO^j(t+s)Vy^2V \cO^j(t-s)y^2] \nonumber \\
&+& 2\mu\sum_j\int_0^tds \  {\rm
	tr}[\cO^j(t+s) V y^2 \cO^j(t-s) V  y^2]. \nonumber
\end{eqnarray}
where $V=V(0)$, $y$ represents the fourth root of the thermal density matrix, Eq.~\eqref{h7}, and the trace is performed with respect to the many-body eigenstates $|n\rangle$ of $H_L$.
Operators $\cO^j$ enter through the coupling $H_I$ which is assumed to have the Maldacena-Qi form  Eq.~\eqref{eq:coupling_Qi}.

The trace on the first line is a time-independent constant, and the trace on the second line is a naturally time-ordered four-point correlator.
Crucially, the trace on the third line has the structure of a regularized  OTOC for all $s$ inside the integration bounds.
Near the lower bound $s\to 0$ it coincides with the regularized OTOC. For $s\neq 0$ the trace represents a more general form of the OTOC dependent on two time variables, $F(t_1,t_2)= {\rm tr}[W(t_1)V y^2 W(t_2) V  y^2]$.
In chaotic systems the latter is commonly assumed to behave according to
\begin{equation}\label{ss3}
F(t_1,t_2) \simeq b(t_1-t_2)e^{\lambda_L(t_1+t_2)/2}
\end{equation}
with $b(t)$ an even function of $t$ and $b(0)$ real positive  \cite{Kitaev2018,Romero2019,Gu2019}.
The trace on the last line of Eq.~\eqref{ss2} is then related to $F(t+s,t-s)$ 
and, upon  integration, becomes proportional 
to $B(t) e^{\lambda_L t}$ with $B(t)= \mu\int_0^tds\  b(2s)$.

If we adopt another common assumption \cite{Maldacena2016} that due to their exponential growth at intermediate times OTOCs tend to dominate over NTOCs, we may conclude that the intermediate-time behavior ($J^{-1} \ll t \ll \mu^{-1}$) of the $LR$ two-point correlator of the coupled theory should be well approximated by 
\begin{equation}\label{s3}
iG_{LR}(t,-t)\simeq A +{B}e^{\lambda_L t},
\end{equation}
where $A$ and $B$ are slowly varying functions of $t$ and may be taken as constants. 

Eq.~\eqref{s3} indicates that, remarkably, the Lyapunov exponent
characteristic of a single chaotic system at inverse temperature
$\beta$ can be extracted by measuring the $LR$ causal two-point
correlator in the ground state of two identical such systems, coupled
through static bilinear terms as in Eq.~\eqref{eq:coupling_Qi}.
We remark that an expression similar to Eq.~\eqref{ss2} can be derived
for the {\em retarded} two-point correlator $G_{LR}^{\rm ret}(t,-t)$
which also contains a dominant OTOC contribution at intermediate
times. Retarded correlators are often more directly related to
measurable quantities, and we will employ them in
Sec.~\ref{sec:SYK} where we provide an explicit numerical
calculation for the example of coupled SYK models. This shows
approximate exponential growth of the retarded two-point correlator
in the appropriate time interval, which lends support to the
conclusions reached in this subsection.

\subsection{Discussion and caveats}
Our main results obtained in this Section can be summarized as follows. We showed that naturally time ordered correlators, such as the one defined in Eq.~\eqref{h4},  evaluated in the TFD state map to out-of-time-ordered correlators with respect to a single system.
This result is generic and requires only that the correlator in question involves operators from both $L$ and $R$ side of the coupled system.
(Details of the correlators however change the regularization structure, that is, the placement of $y^2$ factors in the corresponding OTOCs.)
The transmutation from a two-sided NTOC to a single-sided OTOC can be intuitively understood  as a consequence of the fact that, effectively, time flows in the opposite direction in the two subsystems forming the TFD.
Mathematically this unusual property follows from Eq.\ \eqref{h3}, which shows that a generic two-sided correlator with respect to the TFD state depends on the sum of the temporal arguments for each subsystem, and not on their difference as would normally be the case.

Further, we argued that ordinary Green's functions should also capture the behavior of OTOCs for small values of the coupling $\mu$ between the subsystems and short times $|t-t'| \ll \mu^{-1}$.
Crucially, such two-point correlation functions are in principle much easier to probe in the laboratory, because the measurement can be performed under equilibrium conditions and the protocol does not require varying any system parameters. We shall discuss a specific example of this in Sec.~IV-B.

\section{Application: two coupled SYK models}\label{sec:SYK}

In this Section, we apply the ideas presented above to a concrete model
recently introduced by Maldacena and Qi \cite{Qi2018} that realizes a
quantum-mechanical dual to an eternal traversable wormhole in (1+1)-dimensional anti-de Sitter spacetime (AdS$_2$) by coupling two identical SYK models. This model is convenient for us for several reasons.
First, the SYK model is known to be maximally chaotic: its OTOC
exhibits exponential growth with an exponent that saturates the
universal chaos bound \cite{Maldacena2016,Maldacena2016b}. Second, the
ground state of two such coupled SYK models is well approximated by
the TFD state \cite{Qi2018}, which allows us to apply the machinery developed in the previous Section in a relatively simple setting.
Third, there are several proposals in the literature for experimental realizations of the SYK model and its variants  \cite{Danshita2017,Pikulin2017,Alicea2017,achen2018,Rozali2018}, making it a potentially fruitful platform for laboratory explorations. 

\subsection{The model}

The model introduced by Maldacena and Qi in Ref.~\cite{Qi2018} has the form
\begin{equation} \label{m1}
H = H_L^{\rm SYK} + H_R^{\rm SYK}  + i \mu \sum_j \chi_L^j \chi_R^j
\end{equation}
where $\mu$ is a constant, and $H_\alpha^{\rm SYK} $ with $\alpha=(L,R)$ describe two identical SYK models
\begin{equation}\label{m2}
H_\alpha^{\rm SYK}  = \sum_{i<j<k<l} J_{ijkl} \chi_{\alpha}^i \chi_{\alpha}^j \chi_{\alpha}^k \chi_{\alpha}^l~,
\end{equation}
each involving $N$ Majorana zero-mode operators.  These obey the usual algebra
$\{ \chi_{\alpha}^i , \chi_{\beta}^j \} = \delta^{ij} \delta_{\alpha \beta} ~ , ~ (\chi_{\alpha}^j)^\dagger = \chi_{\alpha}^j$. The coupling constants $J_{ijkl}$ are random, independent Gaussian variables respecting
\begin{equation}\label{m3}
\overline{J_{ijkl}} = 0 ~ , ~ \overline{J_{ijkl}^2} = \frac{3!}{N^3} J^2
\end{equation}
and are \emph{independent} of $\alpha$ -- that is, the disorder in
both SYK models is perfectly correlated. Referring to a potential
experimental realization of the Maldacena-Qi model using quantum dots
(see Sec.~\ref{sec:physical_realizations} below), and for the sake of brevity, we henceforth dub the two SYK subsystems as $L$ and $R$ ``dots''.

 Without loss of generality, we define a complex fermion basis as
\begin{equation}\label{m4}
c_j = \frac{1}{\sqrt{2}}(\chi_L^j - i \chi_R^j)
\end{equation}
to construct the many-body Hilbert space. This basis is helpful because the anti-unitary time-reserval symmetry of the model is manifest and simply represented by $\Theta = \cK$, where $\cK$ denotes complex conjugation. The Majorana operators then transform as $\Theta \chi_L^j \Theta^{-1} = \chi_L^j$ and $\Theta \chi_R^j \Theta^{-1} = - \chi_R^j$. With such a choice $H$ becomes purely real in the many-body basis defined by the operators $c_j$. The coupled SYK system is also invariant under the total fermion parity 
\begin{equation}\label{m5}
P = (-i)^{N} \prod_{j=1}^N \chi_L^j \chi_R^j \cong \left( N_f \right) \; \text{mod} \; 2,
\end{equation}
where $N_f = \sum_j c_j^\dagger c_j$ is the total fermion number. What is less
obvious is that the fermion number modulo 4, $Q_4 \equiv \left(N_f \right) \; \text{mod} \; 4$, is also a symmetry of
$H$. This property relies on the perfectly correlated disorder between the
two subsystems \cite{GarciaGarcia2019}.

The model can be solved in the limit of large $N$ by methods developed in the context of the original SYK model \cite{SY1996,Kitaev2015,Maldacena2016}.
This involves formulating the theory as a Euclidean-space path integral, averaging over the disorder using the replica formalism, and finally writing the large-$N$ saddle-point action for the averaged fermion propagator
$G_{\alpha\beta}(\tau_1,\tau_2)={1\over N}
\sum_j\langle\cT \chi_\alpha^j(\tau_1)\chi_\beta^j(\tau_2)\rangle$.
The resulting saddle-point action reads \cite{Qi2018}
\begin{multline}\label{m6}
S=S_0+{N\over 2}\int_{\tau_1,\tau_2}\sum_{\alpha,\beta}\biggl[
\Sigma_{\alpha\beta}(\tau_1,\tau_2)
G_{\alpha\beta}(\tau_1,\tau_2)\\
-{J^2\over 4} G_{\alpha\beta}(\tau_1,\tau_2)^4\biggr] \\
+i\mu{N\over 2}\int_{\tau_1}[G_{LR}(\tau_1,\tau_1)-G_{RL}(\tau_1,\tau_1)],
\end{multline}
where $S_0=-N\ln{\rm Pf}
(\delta_{\alpha\beta}\partial_\tau-\Sigma_{\alpha\beta})$ and $\Sigma_{\alpha\beta}$
denotes the self energy associated with $G_{\alpha\beta}$.
The corresponding saddle-point equations are obtained by varying the
action with respect to $G_{\alpha\beta}$ and $\Sigma_{\alpha\beta}$.
Using the time-translation invariance, so that
$G_{\alpha\beta}(\tau_1,\tau_2)=G_{\alpha\beta}(\tau_1-\tau_2)$, and
the mirror symmetry between the $L$ and $R$ subsystems, one can write the saddle-point equations in terms of two independent correlators
$G_{LL}(\tau)$ and $G_{LR}(\tau)$. Their frequency-space counterparts are given as
\begin{eqnarray}\label{m7}
G_{LL}(i\omega_n) &=& {i\omega_n -\Sigma_{LL}(i\omega_n)\over
                      D(i\omega_n)}, \\
G_{LR}(i\omega_n) &=&- {i\mu -\Sigma_{LR}(i\omega_n)\over
                      D(i\omega_n)}, \nonumber
\end{eqnarray}
where
$D(i\omega_n)=[i\omega_n-\Sigma_{LL}(i\omega_n)]^2+[i\mu-\Sigma_{LR}(i\omega_n)]^2$
and $\omega_n=\pi T(2n+1)$ is the $n$th Matsubara frequency. The self
energies are given by 
\begin{eqnarray}\label{m8}
\Sigma_{LL}(\tau) &=&J^2 G_{LL}(\tau)^3,\\
\Sigma_{LR}(\tau) &=&J^2 G_{LR}(\tau)^3.
\nonumber
\end{eqnarray}

In the following, we support the ideas for OTOC measurement using the TFD state presented in Sec.\ II  by analyzing numerical solutions of the Maldacena-Qi model defined by Eqs.\ \eqref{m1} and  \eqref{m2}. We perform exact diagonalizations of the Hamiltonian for systems sizes $2N$ as large as $32$, and use the results to calculate various quantities of interest. We also numerically solve the large-$N$ saddle-point equations \eqref{m7} and \eqref{m8} by analytically continuing to real time and frequency domain, and then employing the iterative procedure described in Refs.~\cite{Maldacena2016,Altman2016}.
This yields retarded propagators $G_{\alpha\beta}^{\rm ret}(\omega)$
and their time domain counterparts. Some useful analytical simplifications of the above Schwinger-Dyson (SD) equations and details of our numerical procedures are described in
Appendix \ref{App:numerics}.

\subsection{TFD ground state}

\begin{figure*}[t]
	\includegraphics[width = 0.98\textwidth]{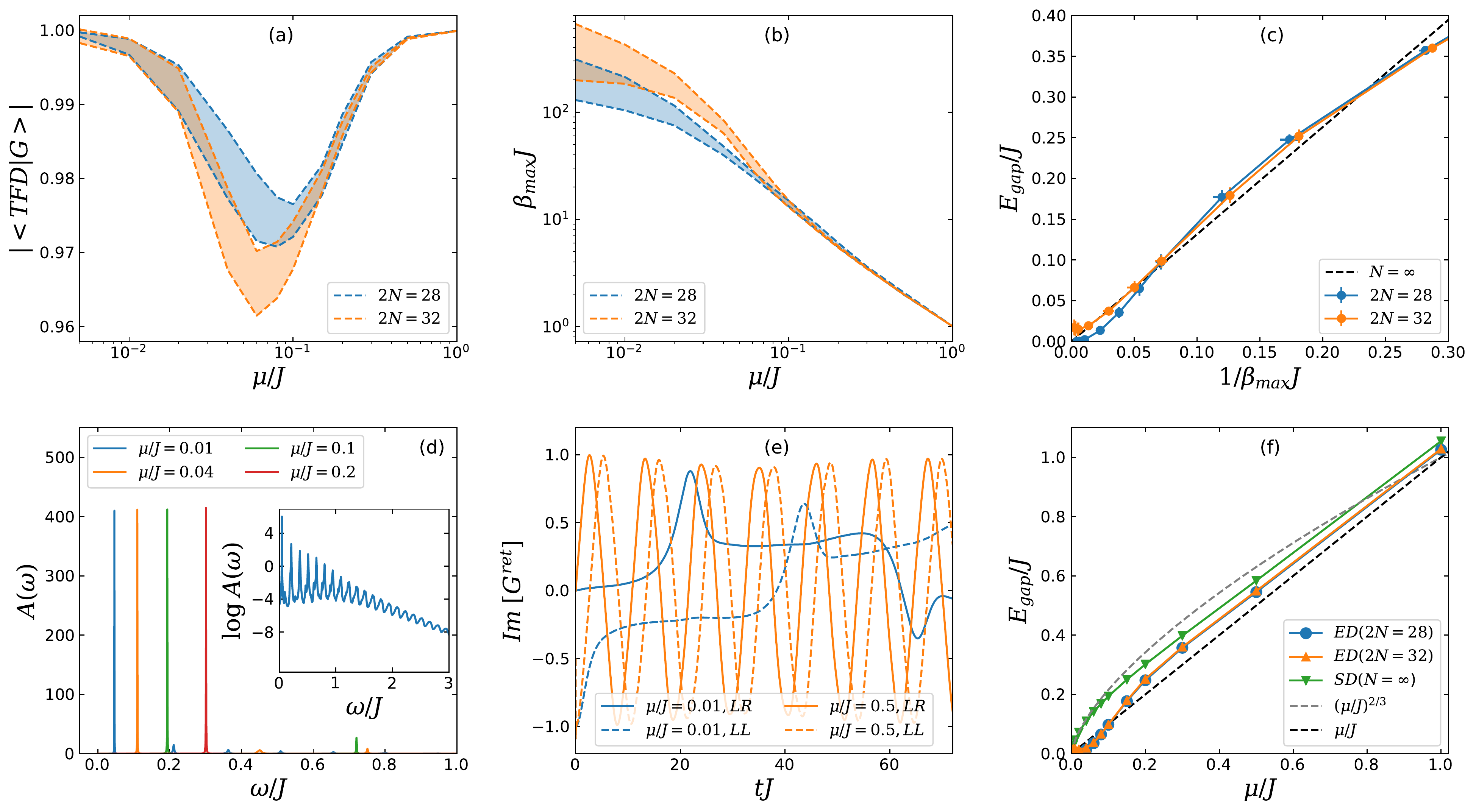}
	\caption{
		Spectral properties of the Maldacena-Qi model obtained numerically through exact diagonalization (ED) for $2N = 28$ and $32$ (a--c), and solving the large-$N$ saddle-point equations (d--e).
		(a) Overlap between the ground state $\ket{G}$ and the best-fit thermofield double state $\ket{\rm TFD_{\beta}}$. The shaded area represents the standard deviation obtained from 16 independent disorder realizations.
		(b) Inverse temperature $\beta_{\rm max}$ characterizing the best-fit $\ket{\rm TFD_{\beta}}$ state.
		(c) Scaling of the energy gap to the first excited state as a function of $1/\beta_{\rm max}J$ in (b). The dashed line indicates the corresponding large-$N$ result obtained in Sec.~\ref{sec:otocs} in the limit of small $\mu/J$.
		(d) Spectral function $A(\omega)$, with the inset showing a series of additional spectral peaks, centered approximately at harmonics $(3n+1)E_\mathrm{gap}$ of the gap.
		(e) Imaginary part of the retarded Green's functions $G^{\rm ret}_{LL}(t)$ and $G^{\rm ret}_{LR}(t)$ in real-time domain.
		(f) Comparison of the energy gaps extracted from ED and large-$N$ saddle-point solution. The expected scalings $E_{\rm gap}\sim (\mu/J)^{2/3}$ and $E_{\rm gap}\sim \mu/J$ at small and large $\mu/J$, respectively, are shown by dashed lines.
		\label{fig2}}
\end{figure*}

In Ref.~\cite{Qi2018} it was argued that the model in Eq.~\eqref{m1} admits an
approximate TFD ground state for all values of the dimensionless
parameter $\mu/J$, and an exact TFD ground state in the limits of
either small or large $\mu/J$. This can be understood intuitively as
follows. For $\mu/J \rightarrow 0$ the ground state of the system is
simply given by $\ket{0}_L \otimes\ket{0}_R$,  which coincides trivially with
the zero-temperature TFD state $\ket{\rm TFD_\infty}$. For $\mu/J \rightarrow \infty$ the
system is best understood as a collection of $N$ decoupled two-level
systems with the Hamiltonian given by the last term in Eq.~\eqref{m1}, and energy levels $\pm \mu$ corresponding to the presence or absence of a fermion in that state. The many-body ground state $\ket{\Psi_0}$ of this system is unique and such that all $N$ single-particle states are empty,
\begin{equation}
c_j \ket{\Psi_0} = 0, ~ \forall j=1\dots N,
\end{equation}
where the $c_j$ are defined in Eq.~\eqref{m4}. This state is equivalent (see Ref.~\cite{GarciaGarcia2019} for an explicit proof) to the infinite-temperature TFD state
\begin{equation}
|{\rm TFD}_0\rangle ={1\over \sqrt{Z_0}}\sum_n |\bar{n}\rangle_L\otimes|n \rangle_R .
\end{equation}
where $\ket{n}$ are the eigenstates of $H^{\text{SYK}}$. For intermediate values of $\mu/J$ one must resort to numerical exact diagonalization, which confirms that the TFD state is always a good approximation to the true ground state of the system, as summarized in Fig.~\ref{fig2}a (see also Refs.~\cite{Qi2018, GarciaGarcia2019}).
The overlap is always greater than $0.96$, with the minimum occurring around $\mu/J \sim 0.1$. This minimum was argued to indicate a phase transition of the Hawking-Page type \cite{Page1983} between a wormhole phase at small $\mu/J$ and low temperature, and a black hole phase at large temperature~\cite{Qi2018,GarciaGarcia2019}.

The parameter $\beta$ characterizing
$\ket{\rm TFD_\beta}$ which best describes the ground state is
monotonically decreasing as a function of $\mu/J$ (see
Fig.~\ref{fig2}b). The energy gap to the first excited state, displayed
in Fig.~\ref{fig2}c, scales as the temperature of the TFD state
$1/\beta$, as expected from the arguments of
Ref.~\cite{Cottrell2019}. In Sec.~\ref{sec:otocs} we obtain the constant of proportionality as $E_{\rm gap} \approx 1.3 T$ from the large-$N$ solution, which agrees well with the ED numerics.
However, as shown in  Fig.~\ref{fig2}f, our ED calculation does not show the scaling
$E_\text{gap} \sim \mu^{2/3} J^{1/3}$ at small $\mu/J$, expected from
the wormhole duality and confirmed by solving the imaginary-time SD equations \eqref{m7} and \eqref{m8} in Ref~\cite{Qi2018}. This is presumably due to finite-size effects which become important at energy scales smaller than $\sim J/N$. 

We can extract the gap amplitude more precisely from the numerical
solution of the large-$N$ saddle-point equations \eqref{m7} and
\eqref{m8}, but now solved in real time and frequency domain.
This is most easily done by analyzing the spectral function
\begin{equation}\label{spec}
A(\omega)=-\frac{1}{\pi}{\rm Im} G_{LL}^\mathrm{ret}(\omega),
\end{equation}
defined using the retarded propagator $G_{LL}^\mathrm{ret}(\omega)$,
which is related to the Matsubara frequency propagator $G_{LL}(i\omega_n )$ by the
standard analytical continuation $i\omega_n\to \omega+i\delta$ \cite{Fetter}. 
The spectral function is  
shown for several values of $\mu$ in Fig.\
\ref{fig2}d. The spectral gap $E_\text{gap}$, defined here as the
position of the first peak in $A(\omega)$, is plotted in
Fig.~\ref{fig2}f. It shows $E_\mathrm{gap} = \mu^{2/3}J^{1/3}$ scaling
(with numerical prefactor very close to 1) for small $\mu/J$, with a crossover to a linear dependence occurring around $\mu/J \approx 0.1$.
An extensive symmetry analysis and substantial simplifications of the
SD equations \eqref{m7}-\eqref{m8}, discussed in Appendix~\ref{App:numerics},
allows us to converge the numerical solution for smaller $\mu/J$ than
was previously reported \cite{Qi2018,GarciaGarcia2019}. This procedure
gives access to the conformal $\sim \mu^{2/3}$ scaling regime and is
also crucial in  providing accurate results for the dynamics of the left-right correlators shown in Fig.~\ref{fig2}e.

\begin{figure*}[t]
	\includegraphics[width=\textwidth]{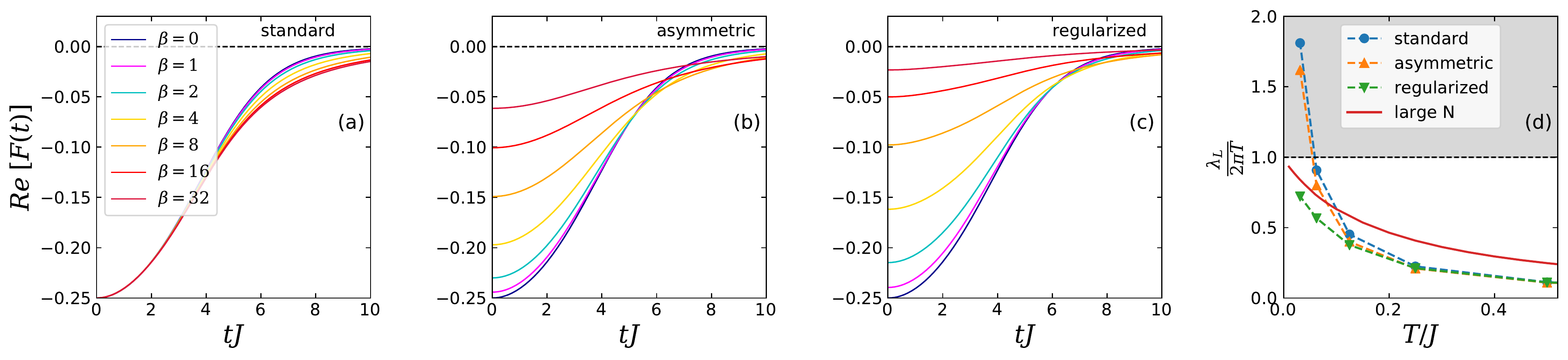}
	\caption{Out-of-time-order correlators $F(t)$ obtained through
		numerical exact diagonalization of a single SYK model with $N=30$
		using  (a) standard, (b) asymmetric and (c) regularized
		forms defined by Eqs.\ \eqref{e3}, \eqref{h10} and
		\eqref{h9}, respectively. The asymmetric and regularized forms of the OTOC correspond to the time-ordered correlators in a TFD state given by Eqs.~\eqref{h10} and \eqref{h4}, respectively. (d) Comparison of the extracted
		Lyapunov exponents $\lambda_L$ for the three different
		regularizations.
		The chaos bound $\lambda_L = 2 \pi T$ is indicated by the horizontal dashed line; the Lyapunov exponent of the SYK model extracted from solving self-consistent ladder diagram equations at large $N$~\cite{Maldacena2016,Altman2016} is shown in red. Taking the latter as a benchmark for finite-$N$ numerical results, we conclude that the regularized OTOC allows to access lower temperatures more reliably than the asymmetric or standard forms.}
	\label{fig3}
\end{figure*}
Note that the spectral function $A(\omega)$ displayed in Fig.~\ref{fig2}d shows intriguing
additional structure, beyond what was reported in previous works. We
find a sharp peak at $\omega=E_{\rm gap}$ followed by an
sequence of peaks centered close to harmonics of
the gap, with spacing $\Delta \omega \sim 3 E_{\rm gap}$. 
The peak at $E_{\rm gap}$ appears to be infinitely sharp (i.e.\
resolution-limited in our numerics), while the harmonics get
progressively broader as shown in the inset of
Fig.~\ref{fig2}d. This structure is reflected in the behavior of
$G^{\rm ret}(t)$ which shows non-decaying oscillations with a period
$2\pi/E_{\rm gap}$ at long times, Fig.~\ref{fig2}e.
The presence of sharp quasiparticle peaks in $A(\omega)$ at low
frequency suggests an emergent Fermi-liquid description at low
energies and temperatures, which is yet to be developed and poses an
interesting challenge for future work.

\subsection{Measuring OTOCs in coupled SYK models}
\label{sec:otocs}

It is known that, in the limit of $N\rightarrow \infty$ and at strong coupling $\beta J \gg 1$, the SYK model is maximally chaotic with a Lyapunov exponent saturating the chaos bound $\lambda_L = 2 \pi T$. In numerical calculations at relatively small $N$, the maximally-chaotic nature of the SYK model, as seen through the Lyapunov exponents, was never reliably observed and the failure was attributed to finite-$N$ effects \cite{Wengbo2016,Pikulin2017}. Indeed, the exponential growth of the OTOC, parametrized by
\begin{equation}
{\rm Re} \left[ F(t) \right] = A + \frac{B}{N} e^{\lambda_L t}
\label{eq:otoc_parametrization}
\end{equation}
with $A$, $B$ real constants, can be expected for times $J^{-1} \lesssim t < 1/\lambda_L \log( N / B)$ and $\beta J < N$.

However, previous numerical calculations were carried out using the standard OTOC in Eq.~\eqref{e3} which shows stronger finite-size effects~\cite{Gu2019,Kobrin2020}.
We compare in Fig.~\ref{fig3} the OTOCs obtained numerically (in a single SYK model) for the three different regularizations discussed in Sec.\ II B above: standard, regularized and asymmetric. We then extract the Lyapunov exponent for each choice by fitting to the expected functional from, Eq.~\eqref{eq:otoc_parametrization}, for intermediate times. Inspired by Ref.~\cite{Shen2017}, we define the lower bound of the fitting region by a time $t_-$ such that $F(t_-) \sim 0.98 F(0)$ which marks the beginning of the exponential growth. Similarly, we define the upper bound $t_+$ as the time at which the second derivative $F''(t_+) < 0$ and thus cannot describe an exponential. For each regularization we observe an exponential growth characteristic of quantum chaotic systems -- however the Lyapunov
exponent $\lambda_L$ extracted from our fitting procedure at low temperature differs drastically between
the three regularizations. Specifically, the standard and asymmetric forms appear to violate
the chaos bound (as also reported elsewhere \cite{Wengbo2016, Pikulin2017}). This is of course not a physical effect, but rather reflects the breakdown of our fitting procedure which occurs because the separation of time scales is insufficient for the small system sizes $N$ considered. 
The regularized form of the OTOC captures the expected trend for the SYK model (red line in Fig.~\ref{fig3}d, cf. Refs.~\cite{Maldacena2016,Altman2016}) more accurately due to weaker finite-size effects (see also Ref~\cite{Kobrin2020}). 

As discussed above, the TFD setup naturally leads to regularized OTOCs
with a square-root of thermal density matrices inserted inside the
trace as indicated in  Eq.~\eqref{h9}.
This is an interesting feature, because such symmetric insertion of
thermal factors does not naturally appear in most other measurement schemes
such as the Lochsmidt echo or those described in Refs.~\cite{Du2017,Rey2017,Monroe2019}.

\begin{figure*}[t]
	\includegraphics[width=\textwidth]{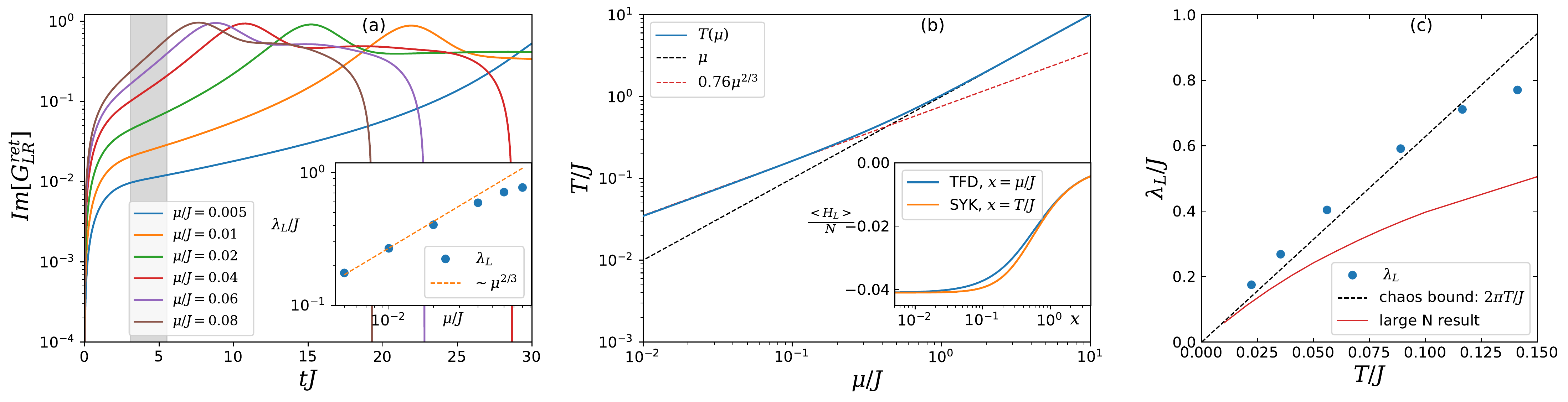}
	\caption{
		Measuring OTOCs using two-point functions in a TFD state.
		(a) Imaginary part of the retarded Green's function $G^{\rm ret}_{LR}(t)$ in real time, obtained from the numerical solution of the saddle-point equations with a small physical temperature $T_{\rm phys}/J \approx 0.001$. At intermediate times $J^{-1} \ll t \ll \mu^{-1}$ an exponential behavior is observed. By fitting the region shown by a shaded area, we extract Lyapunov exponents $\lambda_L(\mu)$ shown in the inset.
		(b) Using the procedure discussed in Sec.~\ref{sec:otocs}, Eqs.~\eqref{hav}-\eqref{hav2}, we obtain the functional dependence between the effective temperature of the TFD state $T=1/\beta$ and the coupling $\mu$. (The inset shows $\langle H_L \rangle$ calculated using both sides of Eq.~\eqref{hav}). This allows to extract $\lambda_L(T)$ in (c) which is consistent with the chaos bound $\lambda_L = 2 \pi T$ at low temperature. The scaling of $\lambda_L(T)$ expected~\cite{Maldacena2016,Altman2016} for the SYK model (as in Fig.~\ref{fig3}) is also shown.
\label{fig4}}
\end{figure*}

As discussed in Sec.~\ref{sec:OTOC_TFD}-D, a possibly more convenient
way to access the OTOC is  through a measurement of the two-sided
Green's function $G_{LR}(t-t')$ in the ground state of the coupled
system. This is clearly a more straightforward
measurement, but is limited to weak couplings $\mu/J$. To verify
that this approach indeed works we adapt Eq.\ \eqref{ss2} to the
Maldacena-Qi model by identifying $\cO^j=\chi^j$. Following the steps
outlined in Sec.~\ref{sec:OTOC_TFD}-D, we derive the short-time expansion
of the retarded version of the averaged $LR$ Majorana propagator, 
\begin{multline}\label{ss2ret} 
iG_{LR}^{\rm ret}(t,-t) = {\theta(t)\over N}\sum_j \langle \{ \chi^j_L(t), \chi^j_R(-t) \} \rangle \simeq\\
+ {4\mu\over N}\sum_{j,k}\int_0^tds \ {\rm Re \ 
	tr}[\chi^j(t+s)\chi^ky^2\chi^k \chi^j(t-s)y^2]  \\
+ {4\mu\over N}\sum_{j,k}\int_0^tds \  {\rm Re \
	tr}[\chi^j(t+s) \chi^k y^2 \chi^j(t-s) \chi^k  y^2],
\end{multline}
valid for $ 0 < t \ll \mu^{-1}$. Similar to the time-ordered case, $G_{LR}^{\rm ret}(t)$ contains an OTOC contribution (last line of Eq.~\eqref{ss2ret}), and we therefore expect an
exponential growth at intermediate times.

In Fig.~\ref{fig4}a we show the imaginary part of $G^{\rm
  ret}_{LR}(t)$ calculated numerically from the large-$N$ saddle-point
equations for several values of $\mu/J$. For sufficiently weak
couplings $\mu/J$, we can fit an approximately exponential growth in
the expected regime, from which we
extract a putative Lyapunov exponent $\lambda_L(\mu)$ as shown in the inset of Fig.~\ref{fig4}a. 
The extracted exponents follow the $\sim \mu^{2/3}$ scaling of the energy gap (see
Fig.~\ref{fig2}f) at small $\mu/J$. Given that $E_{\rm gap}$ scales linearly with the effective temperature  $1/\beta$ of the corresponding TFD state \cite{Cottrell2019} (see
Fig.~\ref{fig2}c), our results imply that $\lambda_L \sim T$, consistent with the expectation for the SYK model at low temperatures.

In order to make quantitative statements, we need to establish the coefficient of proportionality of $\lambda_L(T)$ which requires the knowledge of the function $T(\mu)$. This can be in principle  obtained from our ED results shown in Fig.~\ref{fig2}b.  However, because  ED does not
accurately capture the $E_{\rm gap}\sim\mu^{2/3}$ scaling at small
$\mu/J$, we do not expect this approach to be quantitatively reliable. On the
other hand, as we show below, it is
possible  to extract  the $T(\mu)$ dependence directly from the
large-$N$ formalism which correctly captures the $E_{\rm gap}\sim\mu^{2/3}$ scaling.
To do this we apply Eq.\ \eqref{h1a} with
$\cO_L=H_L^{\rm SYK}$  to the Maldacena-Qi model, obtaining
\begin{equation}\label{hav}
\langle H_L^{\rm  SYK}\rangle_{\rm TFD}=\langle H^{\rm
  SYK}\rangle_\beta.
\end{equation}
The left-hand side is evaluated in the ground state
of the Maldacena-Qi model and gives $\langle H_L\rangle_{\rm TFD}$ as a function of $\mu$ (dropping the SYK superscript from here on). The right hand side
is evaluated in the thermal ensemble of a single SYK model and gives
$\langle H\rangle_\beta$ as a function of temperature. Matching
these two energies through  Eq.~\eqref{hav} then yields the required function $T(\mu)$. 

The expectation value of the Hamiltonian operator can be extracted
from the system Green's functions obtained 
from the large-$N$ saddle point equations.
A textbook procedure \cite{Fetter}  applied to the Maldacena-Qi
Hamiltonian yields
\begin{equation}\label{hav1}
\langle H_L \rangle_{\rm TFD}={N\over
  4}\lim_{\tau'\to\tau^+}\left[ {\partial\over\partial \tau}G_{LL}(\tau'-\tau)+i\mu G_{LR}(\tau'-\tau)\right]
\end{equation}
where $G_{\alpha\beta}(\tau)$ is the imaginary-time Green's
function. Fourier transforming into the Matsubara frequency space and
using the spectral representation of $G_{\alpha\beta}(i\omega_n)$  this can be rewritten
in the integral form 
\begin{equation}\label{hav2}
\langle H_L\rangle_{\rm TFD}={N\over
  4}\int_{-\infty}^\infty d\omega \ n(\omega)\left[\omega\rho_{LL}(\omega)+\mu\rho_{LR}(\omega)\right]
\end{equation}
which is convenient for numerical evaluation. Here
$n(\omega)=1/(e^{\beta\omega}+1)$ denotes the Fermi-Dirac distribution and
  $\rho_{\alpha\beta}(\omega)$ are the spectral functions defined in
  Appendix \ref{App:numerics}. 

We use Eq.\ \eqref{hav2} to evaluate the left-hand side of Eq.\ \eqref{hav}. The right-hand side can be
  obtained in an analogous manner and is given by
  Eq.~\eqref{hav2} with $\mu=0$ and $\rho_{LL}$ replaced by the spectral
  function $\rho(\omega)$ of the single SYK model. The results of this
  calculation are summarized in Fig.~\ref{fig4}b. We find that $T=\mu$
  for large $\mu/J$ while $T/J \approx 0.76 (\mu/J)^{2/3}$ at small $\mu/J$.
 Using this result we obtain Lyapunov exponents in good agreement with the prediction for maximal chaos at low temperatures, $\lambda_L = 2\pi T$, as shown in Fig.~\ref{fig4}c. However, we do not obtain a quantitative agreement with the full solution for the Lyapunov exponents of the SYK model (also shown in Fig.~\ref{fig3}). Our method overestimates the value of $\lambda_L$ for intermediate $T/J$, which could be due to uncertainties in selecting the optimal time window for the fitting procedure, or contamination from the NTOC terms and higher-order terms in $\mu$ in the expansion leading to Eq.~\eqref{ss2ret}.

\section{Physical realizations, measurement schemes}
\label{sec:physical_realizations}

The general scheme to probe quantum chaos using the thermofield double state discussed in Sec.~II is applicable to physical systems of essentially any type. The key requirement is to have two identical copies of the system which can be initialized into the TFD state and then measured.
%
%
Although there are other known systems that exhibit quantum chaos, we continue focusing here on the SYK family of models which are exactly solvable in the large-$N$ limit and have been widely studied in the literature.
Below we discuss possible physical realizations of two coupled SYK models, as well as protocols that yield out-of-time ordered correlation functions by performing legitimate causal measurements. 

\begin{figure}[t]
	\includegraphics[width = \columnwidth]{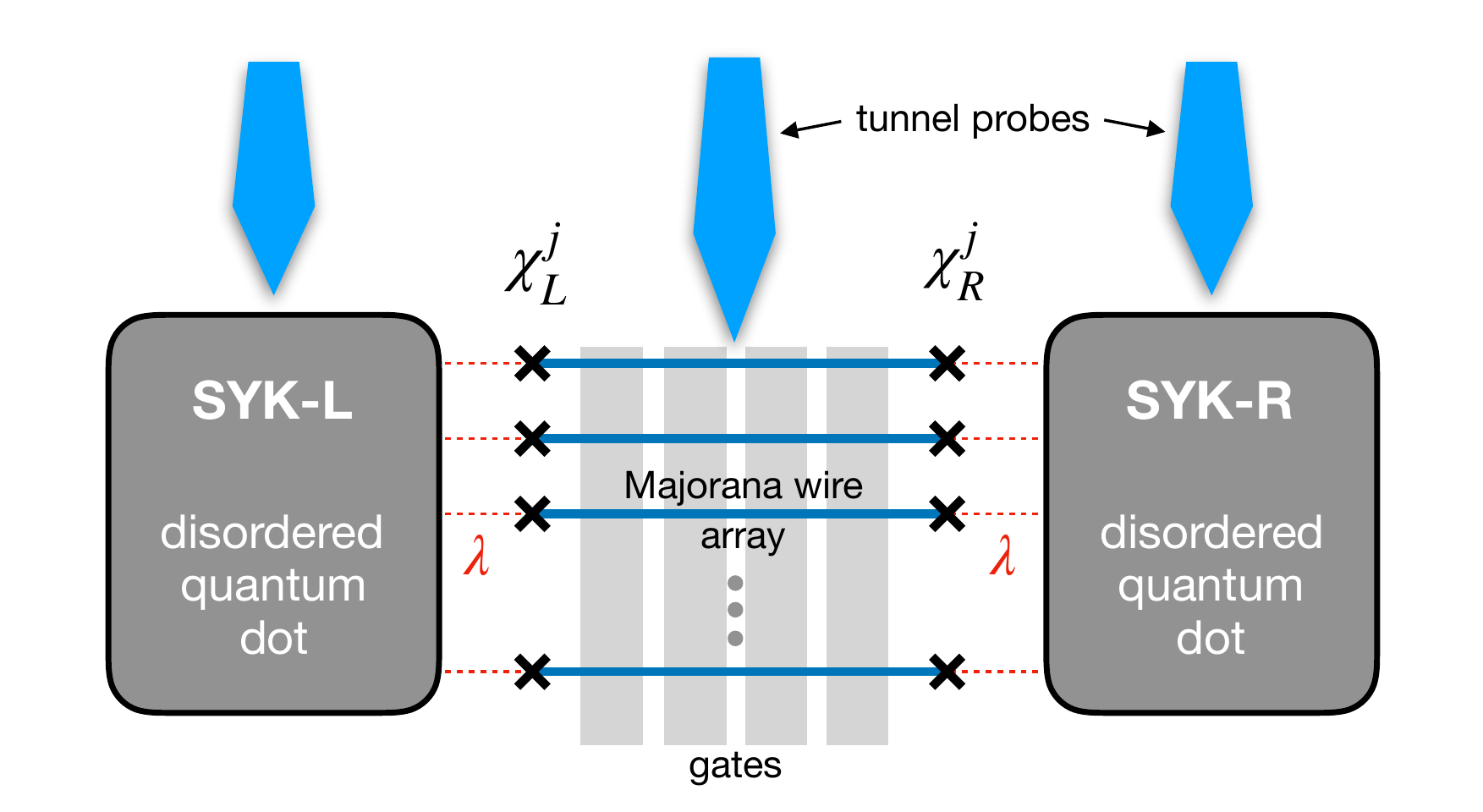}
	\caption{
		Possible realization of the Maldacena-Qi model \cite{Qi2018}
		based on the SYK platform proposed in Ref.~\cite{Alicea2017}. Majorana fermions localized at the ends of
		quantum wires are weakly coupled to two identical quantum dots
		containing electronic disorder.
		Under suitable conditions each quantum dot realizes an SYK model.
		If the wires are relatively short, then overlap between the Majorana wavefunctions in the bulk of each wire leads to their pairwise coupling of the form indicated by the last term in Eq.\ \eqref{m1}. Similar terms arise if capacitive effects are included for the nanowires, cf. Appendix~\ref{AppB}. In both cases, the strength of this coupling can be controlled by the electrostatic gates. Tunnel probes are used to perturb the system by injecting electrons, and allow to perform spectroscopic measurements.    
	}\label{fig:Majoranawires}
\end{figure}

\subsection{Realizations of coupled SYK models} 

\subsubsection{Quantum dots}
Perhaps the most conceptually transparent realization of the Maldacena-Qi model \cite{Qi2018} is depicted in Fig.\ \ref{fig:Majoranawires}.
It consist of $N$ semiconductor quantum wires proximitized to realize a
topological superconductor phase with a pair of Majorana zero modes
bound to their ends \cite{Alicea2012,Beenakker2012,Leijnse2012,Stanescu2013,Elliott2015}.     
The wires are weakly coupled to a pair of identical quantum dots, such
that the zero modes delocalize into them and form two identical SYK
models when interactions between the underlying electrons are taken
into account \cite{Alicea2017}. In a wire of finite length $L$, the
two Majorana endmodes are weakly coupled due to the overlap of their
exponentially decaying wavefunctions in the bulk of the wire.
For the $j$th wire  this coupling has the form $i\mu\chi_L^j\chi_R^j$ with $\mu\sim
e^{-L/\xi}\cos{(k_FL)}$, where $\xi$ denotes the superconducting coherence
length and $k_F$ is the Fermi wavevector of electrons in the
wire. Both $\xi$ and $k_F$ are sensitive to the gate voltage
applied to the wire, which makes the coupling strength $\mu$ tunable,
at least in principle. A similar term arises for long wires $L \gg \xi$ upon including capacitive effects in each quantum wire.
Since the device in Fig.~\ref{fig:Majoranawires} serves as an instructive example below, we expand on some technical details of its realization, following Ref.~\cite{Alicea2017}, in Appendix \ref{AppB}.

While the device depicted in Fig.~\ref{fig:Majoranawires} may look straightforward, its experimental realization presents a significant challenge for reasons that we now discuss. On the positive side there now exists compelling experimental evidence for Majorana zero modes in individual proximitized InAs and InSb wires.
The initial pioneering study by the Delft group \cite{Mourik2012} has been confirmed and extended by several other groups \cite{Das2012,Deng2012,Rokhinson2012,Finck2013,Deng2016,Kouwenhoven2018}.
Assembling and controlling large collections of such wires, as would
be needed in the implementation of a single SYK model, represents a
significant engineering challenge. 
Constructing an identical pair of SYK models entails another level of difficulty.
In the proposal of Ref.~\cite{Alicea2017}, random structure of the SYK coupling constants $J_{ijkl}$ originates from  microscopic disorder that is present in the quantum dot. It is clearly impossible to create two quantum dots that would have identical configurations of microscopic disorder.
A possible solution to this problem would be to engineer quantum dots that are nearly disorder-free, and then introduce strong disorder by hand in a controlled and reproducible fashion. This could be achieved, e.g., by creating a rough boundary or implanting scattering centers in the dot's interior.
In such a situation the electron scattering (and therefore the structure of $J_{ijkl}$) would be dominated by the artificially introduced defects, and two nearly identical quantum dots could conceivably be produced.

\subsubsection{Fu-Kane superconductor}

\begin{figure}[t]
	\includegraphics[width = 7.5cm]{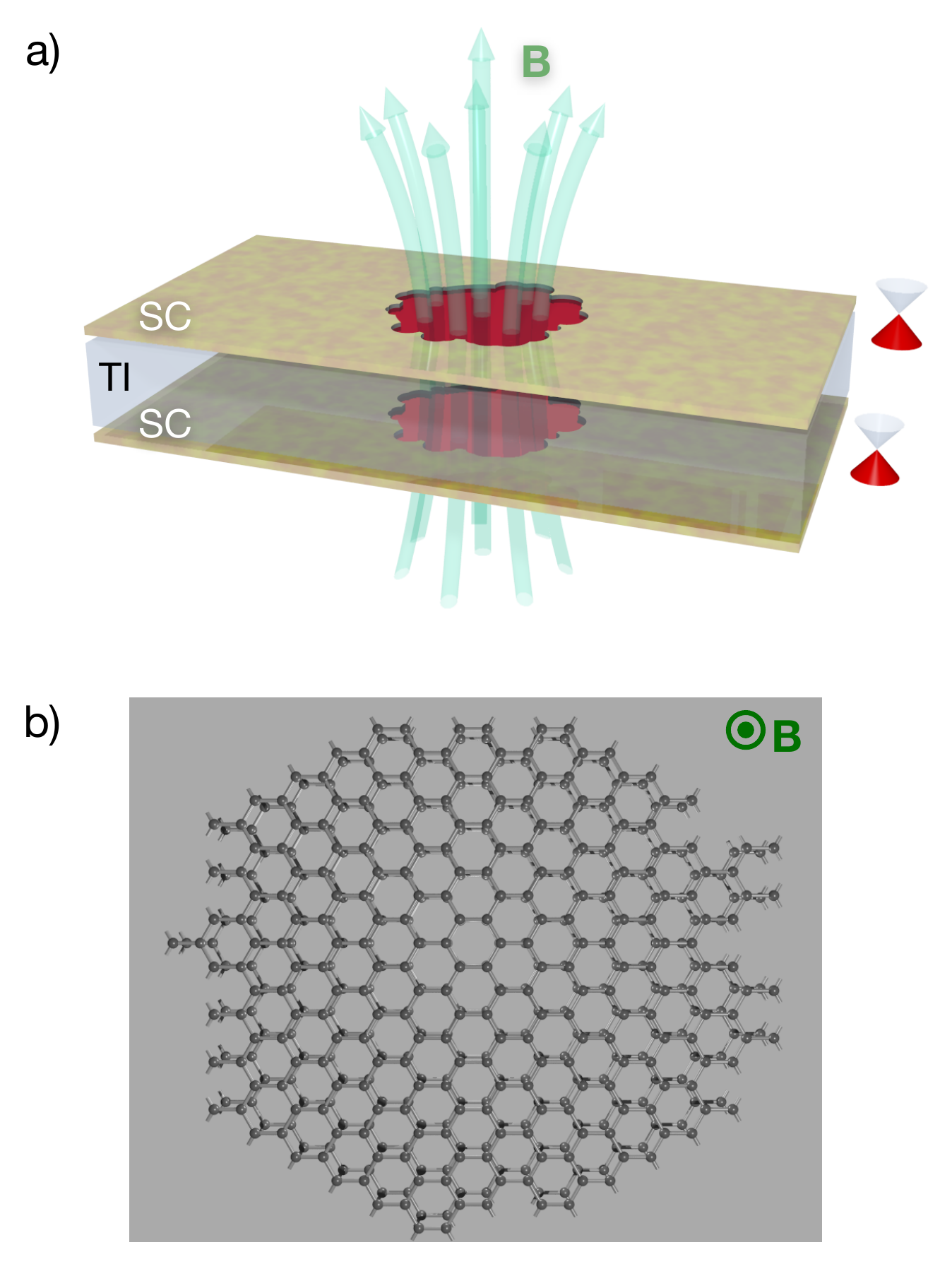}
	\caption{
		Other possible realizations of coupled SYK systems.
		a) A topological insulator film covered by a superconducting film on both sides realizes two copies of the Fu-Kane superconductor. Two identical holes prepared in the SC films, threaded by $N$ flux quanta, then realize two identical weakly coupled SYK models with $N$ Majorana fermions each. b) A bilayer graphene flake with irregular shape in a perpendicular magnetic field $B$ realizes two coupled copies of the cSYK model. 
\label{fig:realizations}}
\end{figure}

Another proposal to realize the SYK model starts from Majorana zero modes localized in vortices at the interface between a topological insulator (TI) and an ordinary superconductor (SC) \cite{FuKane2008}. Specifically, when $N$ such vortices are trapped in a hole fabricated in the superconducting layer, and when the chemical potential of the TI surface state is tuned close to the Dirac point, the effective low-energy description of the system is given by the SYK Hamiltonian \cite{Pikulin2017}.
The random structure of $J_{ijkl}$ here comes from the randomly shaped hole boundary, and can be well-approximated by a Gaussian distribution in certain limits \cite{Pikulin2017, LH2018}. This setup can be turned into a realization of the Maldacena-Qi model, by using a thin film of a TI with a SC layer equipped with an identical hole on each surface, as illustrated in Fig.~\ref{fig:realizations}a.
For a thick film this setup  generates two decoupled identical copies of
the SYK model. For a thin film (e.g. composed of several quintuple layers of Bi$_2$Se$_3$), the tails of Majorana wavefunctions extending into the bulk from the two surfaces will begin to overlap.
The leading term describing such an overlap will be of the form $i\mu\sum_j\chi_L^j\chi_R^j$, as required for the Maldacena-Qi model. The coupling strength $\mu$ here will depend exponentially on the film thickness $d$, but cannot be easily tuned once the device is assembled.

The advantage of this proposal over the quantum dots in Fig.~\ref{fig:Majoranawires} is that randomness in $J_{ijkl}$ here comes from the shape of the hole and is, therefore, under experimental control. Two nearly identical SYK models can conceivably be fabricated in this setup.
On the other hand the experimental status of Majorana zero modes in the Fu-Kane superconductor is not nearly as well developed as in quantum wires.
Experimental signatures consistent with zero modes bound to individual vortices have been reported in ${\mathrm{Bi}}_{2}{\mathrm{Te}}_{3}/{\mathrm{NbSe}}_{2}$ heterostructures \cite{Jia2016a,Jia2016b}, but this result remains unconfirmed by other groups.
More recently signatures  of Majorana zero modes have been observed in surfaces of the iron based superconductor FeTe$_{0.55}$Se$_{0.45}$ \cite{Wang2018,Chen:2018aa,Feng2018,Hanaguri2018,Chiu2019}. 
It is thought that this material is a topological insulator in its normal state, and its surfaces realize the Fu-Kane model  when the bulk enters the superconducting phase below the critical temperature $T_c\simeq 14$K. 

These experimental developments identify the Fu-Kane
superconductor as a promising platform for Majorana device
engineering. Future efforts might bring us closer to realizing the SYK
and Maldacena-Qi models.

\subsubsection{Graphene flake bilayers}

The complex fermion version of the SYK model, sometimes abbreviated as cSYK, exhibits properties in many ways similar to the canonical SYK model with Majorana fermions \cite{SY1996,Sachdev2015}. It is defined by the Hamiltonian
\begin{equation}\label{csyk}
H^{\rm cSYK}=\sum_{ij;kl} J_{ij;kl}c^\dagger_ic^\dagger_jc_kc_l -\tilde{\mu}\sum_jc^\dagger_jc_j,
\end{equation}
where $c_j$ annihilates a complex fermion and $\tilde{\mu}$ is the chemical potential. A realization of the cSYK model has been proposed using electrons in the lowest Landau level of a nanoscale  graphene flake with an irregular boundary
\cite{achen2018}. Once again randomness in $J_{ij;kl}$ originates from the irregular boundary of the flake.

Two identical flakes forming a bilayer illustrated in Fig.\ \ref{fig:realizations}b could realize a complex fermion version of the Maldacena-Qi model if electrons were permitted to tunnel, with weak tunneling amplitude, between the adjacent sites of the two flakes. The tunneling amplitude would depend sensitively on the distance $d$ between the flakes (or on the number of layers in a multi-layer graphene sandwich), but again cannot be easily changed once the device is assembled.
This proposed setup eliminates the need for Majorana zero modes, which is a significant potential advantage. On the other hand, the detailed theory of a TFD-like state and its relation to the ground state of the coupled system has not been worked out for the complex fermion version of the model, and we leave this as an interesting problem for future study.

\subsection{OTOC measurement schemes}
\label{sec:measurement_schemes}

Because the available experimental probes will depend on the specific details of the physical realization, we offer here only general remarks on how OTOC may be measured using the protocols developed in Sec.~II.
For concreteness and simplicity, we focus again on the proposed coupled SYK dot realization of the Maldacena-Qi model, depicted in Fig.~\ref{fig:Majoranawires}, but we expect our discussion to be valid more generally.

At the highest level, we may distinguish two types of situations when attempting to probe OTOC through a causal (time-ordered) measurement: we either have the ability to control the coupling strength $\mu$ on microscopic timescales (i.e.\ times of order $\hbar/J$), or we do not.
In the first case we can manipulate $\mu$ to prepare the initial resource state ${|\rm TFD}_\beta(-t)\rangle$, and then perform a two-sided time-ordered measurement as discussed in Sec.~\ref{sec:OTOC_TFD} and Appendix~\ref{App:negativetimes}. This has the advantage of directly probing the regularized or asymmetric OTOCs. We give some concrete examples of this below. 
If $\mu$ cannot be controlled on microscopic timescales, it is still
possible to extract the OTOC by measuring $G_{LR}^{\mathrm{ret}}(t)$ in a system with
constant nonzero $\mu$, as discussed in
Sec.~\ref{sec:OTOC_TFD}-D. While this measurement is in principle
easier, the quantitative interpretation is less clean, because it
necessitates disentangling of the OTOC  contributions from the NTOC
terms in Eq.~\eqref{ss2}.

\subsubsection{When $\mu$ can be controlled}

Following existing theoretical work on eternal traversable wormholes \cite{Qi2018,Cottrell2019}, we discussed a method to reliably create a TFD state by cooling down to the ground state of a weakly coupled two-system Hamiltonian $H(\lambda)=H_0+\lambda H_I$. Measuring the OTOC requires the TFD state evolved to negative time, $|{\rm TFD}_\beta(-t)\rangle$, which we demonstrate in Appendix~\ref{App:negativetimes} can be achieved, for short time durations at least, by tuning the dimensionless coupling $\lambda$.
We emphasize that in a generic many-body system, this should be a much
easier task than true backward time evolution of a many-body excited
state that would normally be required to measure an OTOC. Such backward
time evolution necessitates the reversal of the sign of the many-body Hamiltonian $H_0$ which is, in the vast majority of cases, not feasible by any known technique.
On the other hand, manipulating the strength of couplings between two systems can often be achieved, e.g., by gating, as discussed in the previous Section and Appendix~\ref{AppB} for the setup of Fig.~\ref{fig:Majoranawires}.

With the above caveats, the proposed protocol to measure OTOC using
the TFD state could be defined follows.
(i) Prepare two identical copies of the system that are weakly coupled and described by $H(1)=H_S$.
(ii) Cool the coupled system to a physical temperature $T_\text{phys}$ that is much smaller than the energy gap of the combined system, which puts it into its ground state well approximated by $|{\rm TFD}_\beta(0)\rangle$.
(iii) Increase coupling $\lambda$ to a value larger than 1 for a short period of time. This creates a good approximation of $|{\rm TFD}_\beta(-t)\rangle$ where the time evolution is with respect to $H_0$, cf. Appendix~\ref{App:negativetimes}.
(iv) Decouple the system by setting $\lambda=0$, and probe it by a conventional measurement. One possibility, mathematically expressed in Eq.~\eqref{h10}, is to excite the system on one side at time $-t$ and then preform a two-sided measurement at time zero. This procedure yields a direct measure of the asymmetrically-regularized OTOC shown in Fig.~\ref{fig3}.

A way to measure two-sided two-body Majorana operators is via wire-charge measurements, cf. Refs.~\cite{Plugge2017,Karzig2017} and detailed in Appendix~\ref{AppB}, that only rely on a capacitive coupling between an external readout circuit and the nanowire charges.
We consider the simplest case where a single, collective gate in Fig.~\ref{fig:Majoranawires} couples to all nanowires. Quantizing a fluctuating global gate voltage $v_g(t) \to [a(t) + a^\dagger(t)]$, and assuming roughly isotropic capacitive coupling parameters $\sim g_j \simeq g$ to all nanowires, one finds
\begin{equation}\label{eq:HresCollective}
H_{\mathrm{charge-readout}} = H_{\mathrm{res}} - g\hat{Q}(t)[a + a^\dagger]~,
\end{equation}
with a single photon species $a$ and total nanowire-charge operator $\hat{Q}(t) = \sum_j \hat{q}_j(t) = \sum_j i\chi_L^j(t)\chi_R^j(t)$. The term $H_\mathrm{res}$ encodes the external resonator readout circuit, and generates the dynamics for resonator photons $a(t)$.
Either the transmission amplitude or - phase shifts in this external circuit, by means of the capacitive coupling in Eq.~\eqref{eq:HresCollective}, then yield a probe of the total nanowire charge $Q(t)$. The latter is directly related to the averaged equal-time left-right Green's function, as $G_{LR}(t) = -iQ(t)/N$.

%
%

\subsubsection{When $\mu$ is fixed}    

In Secs.~\ref{sec:OTOC_TFD}-D and \ref{sec:SYK}-D, we showed that the
averaged $LR$ Green's function of the coupled theory at fixed $\mu$
contains information on the OTOC for small couplings $\mu/J$ and short
times. We now argue that the retarded version of the $LR$   Green's
function [Eq.~(\ref{ss2ret})]   can be probed by a
straightforward tunneling measurement in the setup of
Fig.~\ref{fig:Majoranawires}. Consider a tunnel probe weakly coupled
to one of the wires at a point distance $x$ from its left end  (represented as the central probe in
Fig.~\ref{fig:Majoranawires}). A standard tunneling experiment
measures differential tunneling conductance $g(V)=dI/dV$, which is
proportional to the {\em electron} spectral function in the wire
$\rho_x(\omega)$ at point $x$ and frequency $\omega=eV$, where $V$ is
the applied bias voltage. In Appendix \ref{AppAA} we show that this
quantity is related to the retarded $LR$ two-point Majorana correlator
by a simple relation 
\begin{equation}\label{p4}
iG_{LR}^{\rm ret}(t) \simeq K_x \theta(t)\int_{-\infty}^\infty d\omega
\rho_x(\omega) \sin{\omega t}.
\end{equation}
The constant of proportionality $K_x$ depends on the tunneling matrix
element and on the position $x$ in the wire, but is time independent as the measurement is performed under equilibrium conditions. Therefore, time dependence of $G_{LR}^{\rm ret}(t)$ and the relevant Lyapunov exponent can be extracted from the measured spectral function, using Eq.~\eqref{p4}.

The result given in Eq.\ \eqref{p4} relies on two simple observations, discussed in more detail in Appendix \ref{AppAA}.
First, a retarded time-domain correlator of Hermitian operators is purely imaginary. This fact follows directly from its definition
and underlies the proportionality of $iG_{LR}^{\rm ret}(t)$ to a real quantity.
Second, Majorana zero mode operators $\chi^k_{L/R}$ are simply related
to the electron operators in the wire through 
the solution of the relevant Bogoliubov-de Gennes
equation~\cite{Alicea2012,Beenakker2012,Leijnse2012,Stanescu2013,Elliott2015}, 
which is largely dictated by symmetries of the setup. This implies proportionality of $iG_{LR}^{\rm ret}(t)$ to the electron spectral function, Fourier-transformed into the time domain. 

Based on these observations we expect Eq.~\eqref{p4} to be robust and
independent of system details. Remarkably, in conjunction with Eq.\
\eqref{ss2ret} it connects the Lyapunov exponent of an OTOC with the
electron spectral function in a proximitized semiconductor system,
which is routinely measured in tunneling and other spectroscopic
experiments. 

\section{Conclusions and Outlook}
\label{sec:outlook}

In this work, we introduced and extensively tested the concept of entanglement
in the thermofield double state as a tool to measure
out-of-time ordered correlators in quantum many-body
systems. OTOCs have been of great interest recently because their
exponential growth at intermediate times 
provides direct access to diagnosing quantum-chaotic behavior in many-body
systems.
 
While previous work has implemented OTOC measurements in small-scale
and highly controllable quantum
systems~\cite{Rey2017,Du2017,Monroe2019,Swingle2018}, these approaches
do not lend themselves to the analysis of large, complex many-body
systems that realize quantum chaos in solid-state platforms.
Based on the thermofield double state, one of the main
workhorses for the theoretical description of black hole and wormhole quantum
physics \cite{Maldacena2016b,Gao2017,Yang2017}, we proposed and tested
new protocols for OTOC measurement, where the preparation of a specific resource state -- namely the TFD -- replaces the need for complicated time-evolution or echo procedures at a later stage.
The TFD entangled pair can be obtained as a unique ground state of two coupled
copies of the interacting quantum system under investigation
\cite{Qi2018,Gao2019,Cottrell2019}. We showed that a conventional
measurement with no or only minimal control of the system parameters
can directly access the so-called \emph{regularized} OTOCs. The latter
have been introduced as mathematical objects in field-theoretical
calculations, because in certain limits they are less singular than the
canonical OTOCs. Regularized OTOCs have recently been argued
to measure quantum chaos more reliably than canonical OTOCs
\cite{Galitsky2018,Romero2019,Kobrin2020}, a result corroborated by our numerical
analysis. However, regularized OTOCs are even more difficult to access than canonical OTOCs in a physical measurement, given
that the insertion of square roots of the density matrix on their Schwinger-Keldysh contours (see Fig.~\ref{fig1}) does not
reflect a sensible thermal measurement, even if backward time evolution is considered possible. To our knowledge, only the interferometric approach of Ref.~\cite{Yao2016arxiv} potentially allows for their extraction. It is
all the more exciting that they arise as naturally accessible
objects in our TFD-based protocols.
 
Perhaps the most surprising outcome of our considerations is the
realization, expressed mathematically in Eqs.~\eqref{ss2} and
\eqref{s3}, that the Lyapunov exponent $\lambda_L$ characterizing
quantum chaos is in fact encoded in the intermediate-time behavior of the ordinary two-point correlator $G_{LR}(t)$.
The latter is measured under equilibrium conditions, between operators
drawn from the two subsystems forming the TFD. We confirmed this
result through a  numerical solution of the large-$N$ saddle point
equations associated with two coupled SYK models. These indeed show
approximate exponential growth of $G_{LR}(t)$  at intermediate
times with a Lyapunov exponent $\lambda_L \approx 2\pi T$
consistent with the presence of maximal chaos, saturating the
Maldacena-Shenker-Stanford bound \cite{Maldacena2016b} $\lambda_L\leq
2\pi T$ in the weak coupling limit $\mu/J \ll 1$.
This finding is significant because in many systems such two-point correlators can be probed without much difficulty by spectroscopic techniques.
For example, in the proposed quantum dot realization of two coupled
SYK models illustrated in Fig.~\ref{fig:Majoranawires}, the retarded
Majorana correlator  $G^{\rm ret}_{LR}(t)$ is found to be proportional
to the Fourier transform of the electron spectral function
$\rho_x(\omega)$ (see Eq.~\eqref{p4}) which is accessible through a
routine tunneling measurement.

Last, in this work we have made substantial progress in understanding the structure of the large-$N$ Schwinger-Dyson equations for the Maldacena-Qi model~\cite{Qi2018} comprised of two coupled SYK models, cf. Sec.~III-A and Appendix~\ref{App:numerics}.
We showed that it becomes possible to describe the full real-time dynamics in terms of a single (retarded) Green's function, the corresponding spectral function and a single self-energy.
Finding an explicit analytical solution for the dynamics of such coupled quantum chaotic systems, at least in certain limiting cases, would clearly be very rewarding.
Specifically, as we argued, it should be possible to
extract the intermediate-time exponential growth of $G_{LR}(t)$ and the corresponding
chaos exponent directly from the large-$N$ saddle point equations.
By contrast, in the single SYK model one has to go beyond the saddle point equations and sum an infinite series of ladder diagrams to evaluate the OTOC \cite{Kitaev2015,Maldacena2016}.

As an outlook, interesting future work includes the detailed
investigation of physical platforms for coupled chaotic quantum
systems, for example, based on the ideas presented in Sec.~IV-A and
Refs.~\cite{Pikulin2017,Alicea2017,achen2018}.
The key challenge here will be to prepare two systems that are nearly identical in that the interaction coupling constants $J_{ijkl}$ are essentially the same in both. 
We discussed several possible approaches to this challenge in Sec.~IV-A, but neither is fully satisfactory. Going forward, the most promising route appears to involve an exact microscopic symmetry that would relate two subsystems.
For instance time-reversal $\Theta$ in a systems of spin-${1\over 2}$ fermions would mandate identical $J_{ijkl}$ (up to a complex conjugation) for the two spin projections.
A closely related attractive research direction is towards realizations of ``wormholes'' in coupled complex-fermion SYK models \cite{SY1996,Sachdev2015}. Use of complex fermions would alleviate the need for Majorana zero-modes as basic ingredients and would reintroduce electron spin as a potentially useful degree of freedom.
While it is not obvious at present how to formulate the corresponding complex-fermion TFD state, guidance can be taken from the pedagogical discussion of Ref.~\cite{Cottrell2019}. 

Further, the generality of the construction in Ref.~\cite{Cottrell2019} suggests that many more interesting physical systems, including in higher spatial dimensions, might lend themselves to an investigation of their chaotic behavior using our proposed method. We hope that our findings will stimulate further developments on both theoretical and experimental fronts, which will eventually lead to practical tools for quantum chaos diagnosis in interacting many-body systems.

\section*{Acknowledgements}

The authors are indebted to J. Alicea, V. Galitski, F. Haehl, A. Kitaev, B. Kobrin, X.L. Qi, C. Li, J. Maldacena, S. Sahoo and B. Swingle for useful and stimulating discussions.
We would like to thank B. Kobrin and coworkers for pointing us to subtleties in the numerical analysis of (regularized) OTOCs in the SYK model \cite{Kobrin2020}.
We thank NSERC and CIFAR for financial support. SP and MF are grateful to KITP for hospitality during the conference and program ``Order from Chaos'', where part of the research was conducted with support of the National Science Foundation under Grant No. NSF PHY-1748958. Further SP is grateful to the Aspen Center for Physics, supported by National Science Foundation grant PHY-1607611, for hospitality during the conference ``Many-Body Quantum Chaos''. 

\bibliography{OTOC}

\appendix

\section{TFD preparation at negative times}
\label{App:negativetimes}

To demonstrate the preparation of an initial TFD state $|\Psi_0(-t)\rangle$ at negative times, we consider a generalization of the coupled systems' Hamiltonian $H_S$ to
\begin{equation}\label{h14}
H(\lambda)=H_L+H_R+\lambda H_I,
\end{equation}
where $\lambda$ is a dimensionless parameter used to control the
strength of coupling between the $L$ and $R$ systems. We have
$H(1)=H_S$ and $H(0)=H_0$.
First assume that the system has been cooled down and is in the ground state $|\Psi_0\rangle$ of $H(1)=H_S$,
\begin{equation}\label{h15}
H_S |\Psi_0\rangle=\epsilon_0 |\Psi_0\rangle. 
\end{equation}
Now imagine we increase $\lambda$ to a value larger than one. The
system will start evolving according to
$|\Psi_\lambda(t)\rangle=e^{-iH(\lambda)t}|\Psi_0\rangle$. Using
$H(\lambda)=\lambda H_S-(\lambda-1)H_0$, which follows from Eq.~\eqref{h14},
this evolution can be rewritten as 
\begin{equation}\label{h16}
|\Psi_\lambda(t)\rangle=e^{-i\lambda H_S t+i(\lambda-1)H_0 t}|\Psi_0\rangle.
\end{equation}
We next employ the Baker-Campbell-Hausdorff formula
$e^A e^B=e^{A+B+{1\over 2}[A,B]+\dots}$ to separate the two
terms in the exponential, and multiply from the right by $e^{-B}$ to obtain 
\begin{equation}\label{h17}
e^A =e^{A+B+{1\over 2}[A,B]+\dots}e^{-B},
\end{equation}
where dots represent higher-order commutators. 
Taking $A=-i\lambda H_S t+i(\lambda-1)H_0 t$ and  $B=i\lambda H_S
t$, Eq.~\eqref{h16}  becomes
\begin{equation}\label{h18}
|\Psi_\lambda(t)\rangle=e^{i(\lambda-1)H_0 t-{1\over
		2}\lambda(\lambda-1)[H_0,H_I]t^2+\dots}e^{-i\lambda H_St}|\Psi_0\rangle.
\end{equation}
Using Eq.~\eqref{h15}, the last exponential evaluates to
$e^{-i\lambda\epsilon_0t}$. For short time durations one can
furthermore neglect the $t^2$ term in the first exponential, which leads to
\begin{equation}\label{h19}
|\Psi_\lambda(t)\rangle\simeq e^{-i\lambda\epsilon_0t}  e^{i(\lambda-1)H_0 t}|\Psi_0\rangle.
\end{equation}
We see that increasing the coupling strength $\lambda$ to a value
larger than 1 has the same effect as evolving the state
$|\Psi_0\rangle$ {\em backward in time} under the decoupled
Hamiltonian $H_0=H_L+H_R$. Manipulating the coupling strength
$\lambda$ can therefore be used to prepare the TFD state at negative times,
and represents a simple alternative to engineering a sign inversion of the complicated, interacting Hamiltonian $H_0$.

Eq.~\eqref{h19} remains true for sufficiently short
times $t$ such that one can neglect the commutator term in the
exponential of Eq.~\eqref{h18}. This can be estimated from the
condition 
\begin{equation}\label{h20}
{1\over
	2}\lambda\langle i[H_0,H_I] \rangle t^2 \ll \langle H_0\rangle t.
\end{equation}
If we assume that the relevant energy scales for $H_0$ and $H_I$ are
$J$ and $\mu$, respectively, then $\langle H_0\rangle\sim J$ and
$\langle i[H_0,H_I] \rangle\sim J\mu$. Eq.\ \eqref{h20} hence yields 
a constraint
\begin{equation}\label{h21}
t\ll (\lambda \mu)^{-1}
\end{equation}
on the time duration over which the evolution backward in time
following Eq.\ \eqref{h19} can be achieved. For weakly coupled systems $(\mu\ll J)$, this constraint gives a sufficient
window to probe OTOCs using the method described in Sec.~\ref{sec:OTOC_TFD}--B.
\begin{figure}[t]
	\includegraphics[width = 8.0 cm]{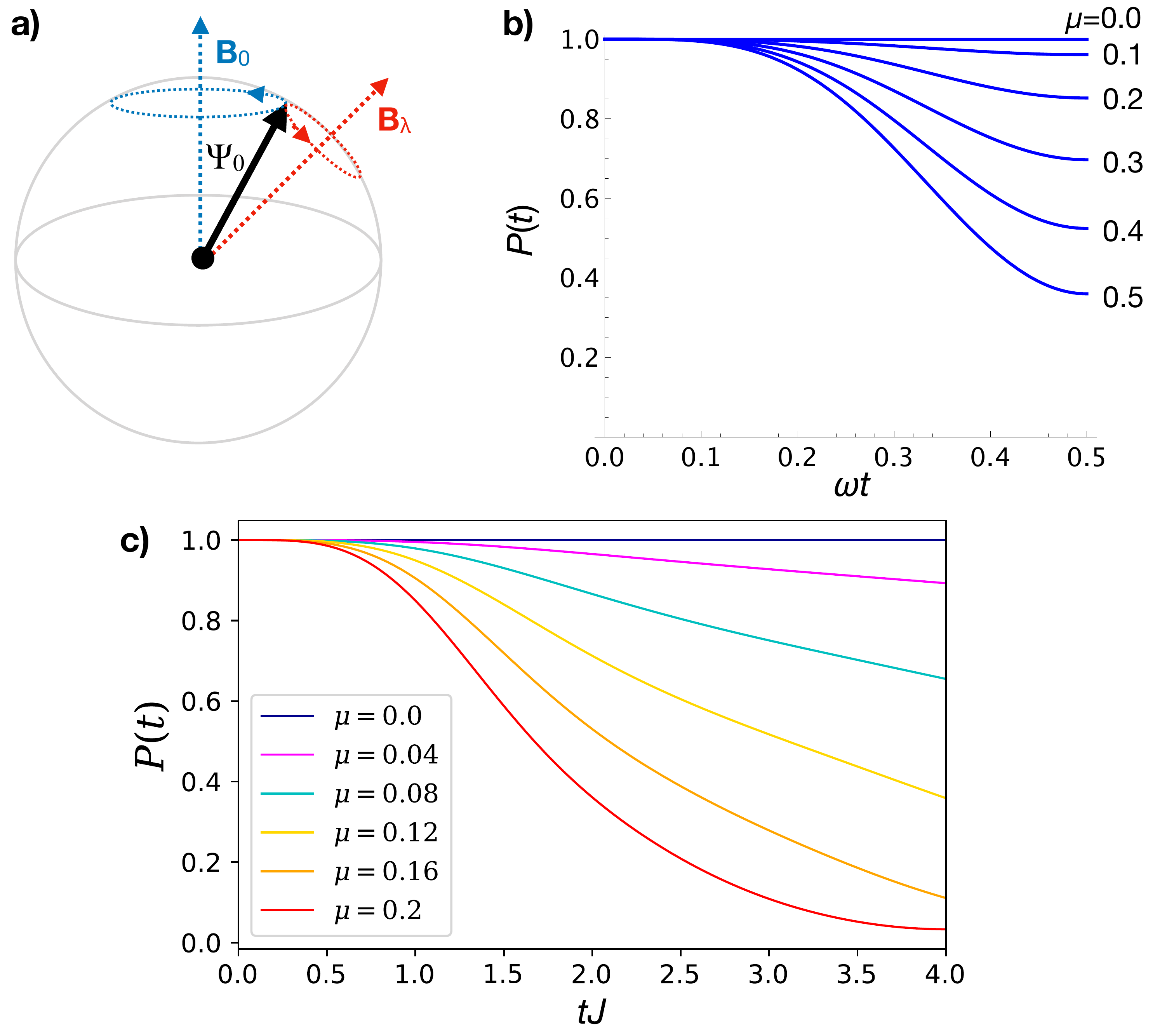}
	\caption{Preparation of the initial state at negative times.
          a) The principle of the backward time evolution illustrated on a simple spin-1/2
		model. The ground state $|\Psi_0\rangle$  of $H(1)$, represented
		by the black arrow on the Bloch sphere, precesses around $\bB_0$
		when $\lambda$ is set to zero. Backward time evolution can be
		approximated, for  short times,  if  $\lambda$ is set to a
		value greater than one and precession around $\bB_\lambda$ occurs.
		b) Overlap $P(t)$ for the spin model. The frequency is defined as
                $\omega=J/2\pi$. c) Overlap $P(t)$ for the
                Maldacena-Qi model calculated using ED with
                $2N=24$. In panels (b) and (c) we use  $J=1$, $\lambda=2$ and several values
		of $\mu$ as indicated.
	}\label{fig:Bloch}
\end{figure}

Some intuitive understanding of the backward time evolution described
above can be gained by analyzing an example of a simple
system. Consider a spin-1/2 degree of freedom in a magnetic field
described by the Hamiltonian $H_0=-J\sigma^z$ and
$H_I=-\mu\sigma^x$. The ground state $|\Psi_0\rangle$ of the combined Hamiltonian
$H(\lambda)=H_0+\lambda H_I$ with $\lambda=1$ has the spin pointing
along the direction parallel to the total magnetic field
$\bB=(\mu,0,J)$. If we switch off $H_I$ the spin will start precessing
counterclockwise around the field direction 
$\bB_0=(0,0,J)$ associated with $H_0$. This is analogous to the TFD state evolving according
to the decoupled Hamiltonian $H_0$ forward in time.  On the other hand,
if we instead increase the value of $\lambda$, the spin will start
precessing counterclockwise around the new field direction
$\bB_\lambda=(\lambda\mu,0,J)$, as illustrated in Fig.\ \ref{fig:Bloch}a.
At short times $t$ we observe that the evolution for $\lambda>1$ approximates backward
time evolution under $H_0$.  This effect can be quantified by
calculating the overlap 
\begin{equation}\label{h22}
P(t)=|\langle \Psi_0(-t) |\Psi_\lambda(t)\rangle|^2
\end{equation}
between  the ground state evolved backward in
time according to $H_0$,
i.e. $|\Psi_0(-t)\rangle=e^{+iH_0t}|\Psi_0\rangle$ and the ground state
evolved forward in time according to $H(\lambda)$,
$|\Psi_\lambda(t)\rangle=e^{-iH(\lambda)t}|\Psi_0\rangle$.
Elementary but somewhat tedious calculation gives an explicit
expression for $P(t)$ that we plot in Fig.~\ref{fig:Bloch}b, for several values of $\mu$. We observe that for short times and $\mu\ll J$ the overlap
remains very close to 1, confirming that the method indeed yields an
excellent approximation to the state $|\Psi_0\rangle$  evolved
backward in time. Crucially, this backward time evolution does not
require reversing the sign of $H_0$, and is achieved solely by
controlling the strength of the $H_I$ perturbation.

In Fig.~\ref{fig:Bloch}c we present numerical evidence supporting this claim
for the coupled Maldacena-Qi model. Similar behavior as described
above is observed at short times in this interacting  many-body system
suggesting that it is generic and thus can be used to prepare the
required initial state $|\Psi_0(-t)\rangle$ in a wide variety of settings.

\section{Short-time expansion of LR two-point correlator}
\label{AppA}

In this Appendix we derive Eq.\ \eqref{ss2}, which we used in the main
text to argue that the $LR$ two-point correlator  $G_{LR}(t,t')$, defined
in Eq.\ \eqref{s1}, contains at short times information on the OTOC. We
proceed by  evaluating the two-point correlator
\begin{equation}\label{as1}
iG_{LR}(t,t')=\langle\Psi_0 |\cT V_L(t)V_R(t')|\Psi_0\rangle.
\end{equation}
Here $|\Psi_0\rangle$ denotes the ground state of the combined system, described
by the Hamiltonian $H=H_L+H_R+H_I$, which we will approximate later on
by $|{\rm TFD}_\beta\rangle$. We work in the Heisenberg picture where
$V_\alpha(t)$ is an arbitrary Hermitian operator evolving according
to the full Hamiltonian $H$. 

$G_{LR}(t,t')$ is a naturally time-ordered correlator that a physical probe would measure in the ground state of the combined system. We would like to know how this quantity is related to what a physical probe would measure in a thermal ensemble at inverse temperature $\beta$ of a {\em single, decoupled} system.
Mathematically, the goal is to express $G(t,t')$ as an average  with
respect to the TFD state  of operators that evolve according to $H_0=H_L+H_R$. To proceed, we pass to the interaction picture by writing \cite{Fetter}
\begin{equation}\label{as2}
V_\alpha(t)=U(0,t)V_\alpha^I(t)U(t,0),
\end{equation}
where superscript $I$ denotes the interaction picture and 
$U(t,t')=e^{iH_0t}e^{-iH(t-t')}e^{-iH_0t'}$ is the unitary operator that
translates between the Heisenberg and interaction pictures.  Using
Eq.\ \eqref{as2}, the correlator becomes
\begin{equation}\label{as3}
iG_{LR}(t,t')=\langle\Psi_0 | U(0,t)
V_L^I(t)U(t,t') V_R^I(t')U(t',0)|\Psi_0\rangle,
\end{equation}
where we used the property $U(t,s)U(s,t')=U(t,t')$. For simplicity, we henceforth also assume that $t>t'$.

We now employ a standard result of diagrammatic many-body theory \cite{Fetter}
that express $U(t,t')$ as a series expansion in $H_I(t)$ of the form
\begin{eqnarray} 
U(t,t')&=&1+(-i)\int_{t'}^t dsH_I(s) \label{as4} \\
&+&{(-i)^2\over 2}\int_{t'}^t ds_1\int_{t'}^t ds_2\cT[H_I(s_1)
H_I(s_2)] +\dots \nonumber 
\end{eqnarray}
Here the time evolution of $H_I(t)$ is according to
$H_0$. Substituting this into Eq.~\eqref{as3}, and retaining only terms
up to first order in $H_I(t)$, we find
\begin{eqnarray} 
iG_{LR}(t,t')&\simeq&\langle V_L^I(t) V_R^I(t')\rangle_0  \label{as5} \\
&-&i\int_t^0ds  \langle H_I(s) V_L^I(t) V_R^I(t')\rangle_0 \nonumber \\
&-&i\int_{0}^{t'}ds  \langle V_L^I(t) V_R^I(t')H_I(s) \rangle_0 \nonumber \\
&-&i\int_{t'}^tds  \langle V_L^I(t) H_I(s)  V_R^I(t')\rangle_0. \nonumber
\end{eqnarray}
where $\langle \dots \rangle_0$ denotes the expectation value with
respect to the ground state $|\Psi_0\rangle$. This expression is
valid when one can neglect all higher order terms in the
expansion \eqref{as4} of $U(t,t')$, $U(0,t)$ and $U(t',0)$. This
requires short time durations $|t-t'|$ as well as
individually small $|t|$ and $|t'|$. In the following we focus on the symmetric case
$t'=-t$, which has a convenient property that short  duration $|t-t'|$
automatically assures that $|t|$ and $|t'|$ are small.

As the final step we substitute for $H_I$ the Maldacena-Qi form given in Eq.\ \eqref{eq:coupling_Qi}, and approximate
$|\Psi_0\rangle$ by $|{\rm TFD}_\beta\rangle$.
With these choices, the expectation values in Eq.~\eqref{as5} can be expressed in terms of single-sided averages using the procedure explained in Sec.~II-B of the main text, see especially the steps leading to Eq.~\eqref{h8}.
The correlator thus becomes
\begin{eqnarray} 
iG_{LR}(t,-t)&\simeq&{\rm tr}[V(-t)y^2 V(-t)y^2] \label{as6} \\
&-&\eta\mu\sum_j\int_0^tds \  {\rm tr}[V(-t)\cO^j(-s)y^2\cO^j(s)V(-t)y^2]  \nonumber \\
&-&\eta\mu\sum_j\int_{-t}^0ds\  {\rm tr}[\cO^j(-s)V(-t)y^2V(-t)\cO^j(s)y^2]  \nonumber \\
&+&\mu\sum_j\int_{-t}^tds\  {\rm tr}[\cO^j(-s) V(-t) y^2\cO^j(s)V(-t)y^2],  \nonumber 
\end{eqnarray}
where $\eta=+/-$ for bosonic/fermionic operators. We dropped superscript $I$ and subscripts $R/L$ on all operators;
it is understood that they now evolve according to the single-sided Hamiltonian, say $H_L$, while the traces are taken with respect to the eigenstates $|n\rangle$ of the same Hamiltonian.
Each individual trace in Eq.~\eqref{as6} is time-translation invariant, and in the following we find it convenient to shift all temporal
arguments of operators inside the traces by $+t$.
In addition, using the cyclic property of the trace, we may combine
the second and third lines and express the last line as an integral from 0 to $t$. This leads to the form quoted in the main text Eq.~\eqref{ss2}.

Mathematically, Eq.\ \eqref{as6} can be viewed as an expansion of the
propagator $G$  in
powers of a dimensionless time variable $\tilde{t}=\mu t$ to first order. Higher
order contributions that would result from the omitted terms in Eq.\
\eqref{as4} can be neglected when $\mu t\ll 1$ which constrains
the expected domain of validity of Eq.\ \eqref{as6} to short times or small
values of coupling $\mu$.

\section{OTOC from electron spectral function}
\label{AppAA}

In this Appendix we derive Eq.\ \eqref{p4}, which provides a simple
route to access OTOC and Lyapunov exponent through an
equilibrium tunneling measurement of the electron spectral function in
a wire that forms a part of the device shown in Fig.~\ref{fig:Majoranawires}.

As a first step, we show that a time-domain retarded propagator of Hermitian operators is imaginary valued. Consider a retarded propagator defined as 
\begin{equation}\label{aa1}
iG^{\rm ret}(t,t')=\theta(t-t')\langle\{A(t),B(t')\}\rangle_0
\end{equation}
evaluated in the ground state $|\Psi_0\rangle$ of the
system. Expanding the anticommutator and using the basic property  of
the inner product $\langle a|b\rangle=\langle b|a\rangle^*$,
we can rewrite the average as 
$\langle A(t)B(t')\rangle_0 +\langle
[B(t')A(t)]^\dag\rangle^\ast_0$. For Hermitian operators $A$ and $B$
this equals $2{\rm Re}\langle A(t)B(t')\rangle_0$, and $G^{\rm ret}(t,t')$ is therefore purely imaginary.

As we already mentioned, a tunneling measurement can be used to extract
the electron spectral function $\rho_x(\omega)$ which is related to the
electron propagator $\cG_x(\omega)$. We therefore start by considering
the corresponding quantity defined in the time domain as
\begin{equation}\label{aa2}
i\cG_x(t)=\langle \cT c_x(t)c_x ^\dag (0) \rangle_0,
\end{equation}
where $c_x(t)$ annihilates electron at time $t$ and at point $x$ of
the wire.  We want to relate $\cG(t)$ to the Majorana propagators
$G_{\alpha\beta}(t)$ discussed in the main text. To this end recall
the relations
\begin{eqnarray}  
\chi_L&=& \int_0^L dx \Phi_L(x)[c^\dag_x+c_x] \label{aa3}\\
\chi_R&=& i\int_0^L dx \Phi_R(x)[c^\dag_x-c_x]  \nonumber
\end{eqnarray}
which follow from the solution of the Bogoliubov-de Gennes (BdG) equations
for the Majorana wire
\cite{Alicea2012,Beenakker2012,Leijnse2012,Stanescu2013,Elliott2015}. Here
$\Phi_{L/R}(x)$ represent the Majorana wavefunctions. They are
real-valued and peaked at the $L$ or $R$ end of the wire,
respectively, with  exponentially decaying tails extending into the
wire. The form of Eqs.\ \eqref{aa3} is constrained by the choice that
$\chi_{L/R}$ transform as even/odd under time reversal. 

We may invert Eqs.\  \eqref{aa3} to express the electron operators in
terms of the Majoranas,
\begin{eqnarray}  
c_x&\simeq& \Phi_L(x)\chi_L+i\Phi_R(x)\chi_R +\dots, \label{aa4}\\
c_x^\dag&\simeq& \Phi_L(x)\chi_L-i\Phi_R(x)\chi_R +\dots,  \nonumber
\end{eqnarray}
where the dots represent the remaining quasiparticle operators that
form the complete set of solutions of the BdG equations. We will
assume that these occur at non-zero energies separated from the
zero-mode manifold by a gap, and will thus not affect the low-energy
spectral function.  Substituting into Eq.\ \eqref{aa2} and neglecting
these terms, we find the Majorana fermion contribution to the electron
Green's function
\begin{multline}  \label{aa5}
\cG_x(t)\simeq\Phi_L^2(x)G_{LL}(t)+\Phi_R^2(x)G_{RR}(t) \\
-i\Phi_L(x)\Phi_R(x)[G_{LR}(t)-G_{RL}(t)]. 
\end{multline}
Making a further non-essential assumption of mirror symmetry between
the $L$ and $R$ sides of the system, we have $G_{LL}(t)=G_{RR}(t)$ and
$G_{LR}(t)=- G_{RL}(t)$. Also, we note that the same calculation can
be repeated for the retarded function $i\cG^{\rm
  ret}_x(t)=\theta(t)\langle \{c_x(t),c_x ^\dag (0)\} \rangle_0$, with
a similar result: 
\begin{multline}  \label{aa6}
\cG_x^{\rm ret}(t)\simeq[\Phi_L^2(x)+\Phi_R^2(x)]G_{LL}^{\rm ret}(t) \\
-2i\Phi_L(x)\Phi_R(x)G_{LR}^{\rm ret}(t). 
\end{multline}
Given that $G_{LL}^{\rm ret}(t)$, $G_{LR}^{\rm ret}(t)$ are imaginary while $\Phi_{L/R}(x)$ are real, we find 
\begin{equation}  \label{aa7}
iG_{LR}^{\rm ret}(t)\simeq - {{\rm Re}[\cG_x^{\rm ret}(t) ]\over 
2\Phi_L(x)\Phi_R(x)}. 
\end{equation}

To complete this calculation, it remains to relate ${\rm Re}[\cG_x^{\rm ret}(t)]$
to the electron spectral function which is the observable quantity. We use the spectral representation $\cG_x^{\rm ret}(\omega)=\int_{-\infty}^\infty d\omega'\rho_x(\omega')/(\omega-\omega'+i\delta)$,
which, upon Fourier transforming, gives
\begin{equation}  \label{aa8}
\cG_x^{\rm ret}(t)=i\theta(t) \int_{-\infty}^\infty
  d\omega e^{-i\omega t} \rho_x(\omega).
\end{equation}
Taking the real part and recalling that $\rho_x(\omega)$ is strictly real, we obtain
\begin{equation}  \label{aa9}
{\rm Re}[\cG_x^{\rm ret}(t)]=\theta(t) \int_{-\infty}^\infty
  d\omega \rho_x(\omega) \sin{\omega t}.
\end{equation}
Finally combining with Eq.~\eqref{aa7}, we arrive at  Eq.~\eqref{p4} of the main text.

\section{Majorana nanowire device}
\label{AppB}

One of our example realizations of two coupled chaotic systems is the two-sided SYK device shown in Fig.~\ref{fig:Majoranawires}, inspired by the SYK setup of Chew, Essin and Alicea~\cite{Alicea2017}. Since this is the most transparent and directly controllable realization we discuss, let us here expand on its technical underpinning in the framework of Ref.~\cite{Alicea2017}.\\
The system in Fig.~\ref{fig:Majoranawires} is described by $N$ sets of Majorana modes $\chi_L^j,~\chi_R^j$ at the left and right ends of the nanowires, and by the $N_{L/R} \gg N$ complex fermions $c_{s,\alpha=L/R}$ hosted in disordered wave-functions of its quantum dots.
We write the dot fermions in Majorana representation $c_{s,\alpha} =
(\eta_{s,\alpha} + i\tilde{\eta}_{s,\alpha})/2$, where
$\eta_{s,\alpha}$ is even under time-reversal (TR) while
$\tilde{\eta}_{s,\alpha}$ is odd. Similarly we take nanowire Majoranas
$\chi_L^j$ and $\chi_R^j$ to be even and odd under TR,
respectively. Assuming that both quantum dots preserve the BDI
symmetry class of the Majorana sector \cite{Alicea2017}, the left and
right ends of the device then are guaranteed to host $N$ TR-even
(TR-odd) Majorana zero modes. This statement holds unless the $LR$ couplings $\sim i\chi_L^j\chi_R^j$ are introduced, where in the main text we discuss the full crossover from weak to strong bilinear couplings.\\
To express the toy-model Hamiltonian describing the device in Fig.~\ref{fig:Majoranawires}, it is now convenient to introduce Majorana spinors $\vec{\chi}_{L/R}$, $\vec{\eta}_{L/R}$ and $\vec{\tilde{\eta}}_{L/R}$ for the respective left/right groups of Majorana fermions. Following Ref.~\cite{Alicea2017}, we first describe the hybridization of Majoranas $\vec{\chi}_{L/R}$ into the left and right quantum dots as
\begin{equation}
H_0 =
i\left(\vec{\chi}_L^T M_{\lambda L} + \vec{\eta}_L^T M_{\epsilon L} \right) \vec{\tilde{\eta}}_L
+ i \vec{\eta}_R^T
\left( M_{\lambda R} \vec{\chi}_R + M_{\epsilon R} \vec{\tilde{\eta}}_R\right) ~.
\end{equation}
Here $M_{\lambda \alpha}$ are real rectangular matrices of couplings $\sim \lambda$ between the wire and dot Majoranas, cf. Fig.~\ref{fig:Majoranawires}, and $M_{\epsilon \alpha}$ are real $N_\alpha \times N_\alpha$ matrices encoding the dots level structure $\sim \epsilon$.
The coupling Hamiltonians $H_{0, \alpha=L/R}$ can be diagonalized by orthogonal rotations obtained from a singular-value decomposition of the coupling matrix encoded in $M_{\lambda \alpha}$ and $M_{\epsilon \alpha}$ \cite{Alicea2017}, giving new Majorana spinors
\begin{equation}\label{eq:OrthoTrans}
\begin{pmatrix}
\vec{\chi}_L' \\
\vec{\eta}_L'
\end{pmatrix} = 
\begin{pmatrix}
\cO_{\chi\chi, L}  & \cO_{\chi\eta, L}\\
\cO_{\eta\chi, L} & \cO_{\eta\eta, L}
\end{pmatrix}
\begin{pmatrix}
\vec{\chi}_L \\
\vec{\eta}_L
\end{pmatrix}
\end{equation}
and $\vec{\tilde{\eta}}_L' = \cO_{\tilde{\eta}, L} \vec{\tilde{\eta}}_L$, and similar for right Majorana operators with $L\to R$ and $\vec{\eta} \leftrightarrow \vec{\tilde{\eta}}$.
In these new operators,
\begin{equation}\label{eq:Hdiagonal}
H_{0, \alpha=L/R} =
i \sum_{j=1}^{N_\alpha} E_{j,\alpha} \eta_{j,\alpha}' \tilde{\eta}_{j,\alpha}'~~,
\end{equation}
where $\eta_{j,\alpha}'$ and $\tilde{\eta}_{j,\alpha}'$ correspond to the rotated spinors $\vec{\eta}_\alpha'$ and $\vec{\tilde{\eta}}_\alpha'$, and $E_{j,\alpha}$ are associated hybridization energies obtained from rotating the coupling matrix.
Note that as guaranteed by the BDI classification, each dots hosts $N$ zero-modes encoded in spinors $\vec{\chi}_\alpha'$ that do not appear in $H_{0,\alpha}$. Further, assuming strong wire-dot hybridizations $\lambda \gg N \delta\epsilon$, where $\delta\epsilon$ is the typical level spacing in the dot, all wire Majoranas $\vec{\chi}_\alpha$ are absorbed into the respective left and right dots. Finite-energy modes in Eq.~\eqref{eq:Hdiagonal} are split off to energies $\sim N \delta\epsilon$ by level repulsion.\\
We now add the $LR$ coupling between the two SYK-dots. In the original basis, sensible couplings are pairwise between Majoranas $\chi_L^j$ and $\chi_R^j$ on each wire, yielding
$H_\mathrm{int} =
i \sum_{j=1}^{N} \mu_j \chi_L^j\chi_R^j = i \vec{\chi}_L^T \mu_{LR}\vec{\chi}_R$
with diagonal coupling matrix $\mu_{LR}$. The specific form and possible tuning of couplings $\mu_j$ is discussed below.
Inverting the orthogonal transformations in Eq.~\eqref{eq:OrthoTrans}, one can represent the original wire Majoranas in the new variables as
$\vec{\chi}_L = \cO_{\chi\chi, L}^T \vec{\chi}_L' + \cO_{\eta\chi, L}^T \vec{\eta}_L'$ and 
$\vec{\chi}_R = \cO_{\chi\chi, R}^T \vec{\chi}_R' + \cO_{\tilde{\eta}\chi, R}^T \vec{ \tilde{\eta}}_R'$
and obtain
\begin{equation}
H_\mathrm{int} =
i \vec{\chi}_L'^T \mu_{\chi\chi} \vec{\chi}_R'
+ i \vec{\chi}_L'^T \mu_{\chi\tilde{\eta}} \vec{\tilde{\eta}}_R'
+ i \vec{\eta}_L'^T \mu_{\eta\chi} \vec{\chi}_R'
+ i \vec{\eta}_L'^T \mu_{\eta\tilde{\eta}} \vec{\tilde{\eta}}_R'
\end{equation}
Here the new $LR$ coupling matrices are given as $\mu_{\chi\chi} = \cO_{\chi\chi, L} \mu_{LR} \cO_{\chi\chi, R}^T$ etc., connecting different species of the new Majorana operators.
Assuming that terms $\sim \mu_j$ are small against wire-dot hybridizations and the on-site level repulsion scale, even in their presence the gapped Majorana-pairs in Eq.~\eqref{eq:Hdiagonal} will stay far-removed. The remaining effective  Hamiltonian then reads
\begin{equation}
H_\mathrm{int, eff} \simeq i \vec{\chi}_L'^T \mu_{\chi\chi} \vec{\chi}_R'~,~~\mu_{\chi\chi} = \cO_{\chi\chi, L}\mu_{LR}\cO_{\chi\chi, R}^T~.
\end{equation}
In the absence of left-right symmetry of the underlying disordered quantum dots in Fig.~\ref{fig:Majoranawires}, there is no reason to assume that $\mu_{\chi\chi}$ is diagonal. Nevertheless, it is always possible to re-diagonalize the $LR$ Hamiltonian by re-applying the orthogonal transformations to Majoranas $\vec{\chi}_{L/R}'$.
Given that the original coupling matrix $\mu_{LR}$ was diagonal, one finds $H_\mathrm{int, eff} = i \vec{\chi}_L''^T \mu_{LR} \vec{\chi}_R''$ with
\begin{equation}
\vec{\chi}_L'' = \cO_{\chi\chi, L}^T \vec{\chi}_L'
= \cO_{\chi\chi, L}^T\cO_{\chi\chi, L} \cdot \vec{\chi}_L + \cO_{\chi\chi, L}^T \cO_{\chi\eta, L} \cdot \vec{\eta}_L~,
\end{equation}
and similarly $\vec{\chi}_R'' = \cO_{\chi\chi, R}^T \vec{\chi}_R'$.
Note $\cO_{\chi\chi, s}^T\cO_{\chi\chi, s} \neq 1$, since only the full transformation in~\eqref{eq:OrthoTrans} is orthogonal.\\
We hence observe that under the approximations taken above, the
interaction matrix $\mu_{LR} = \mathrm{diag}(\mu_j)$ remains unchanged
by the absorption of Majorana modes $\chi_{j,L/R}$ into the SYK dots, but now acts on the new modes $\vec{\chi}_{L/R}''$.

Finally, for ease of notation, we relabel $\vec{\chi}_{L/R}'' \to \vec{\chi}_{L/R}$ and proceed to add on-site interactions for the remaining Majorana zero-modes, inherited from Coulomb interactions of the dot fermions $c_{s, \alpha=L/R}$ \cite{Alicea2017}.
Generically one then obtains a Hamiltonian $H = H_L + H_R + H_\mathrm{int, eff}$ with site-dependent four-Majorana interactions as in Eq.~\eqref{m2}
\begin{equation}\label{eq:Hfull}
H_{\alpha=L/R}^{\rm SYK}  = \sum_{i<j<k<l} J_{ijkl}^\alpha \chi_s^i \chi_s^j \chi_s^k \chi_s^l~~.
\end{equation}
From here, assuming left-right symmetry of the SYK-dots in
Fig.~\ref{fig:Majoranawires} such that $J_{ijkl}^L = J_{ijkl}^R$
identically, one obtains the Maldacena-Qi Hamiltonian~\cite{Qi2018}
in Sec. III.

We now discuss how one may realize (tunable) one-to-one bilinear couplings across nanowires of the device in Fig.~\ref{fig:Majoranawires}. While one may use residual Majorana hybridizations $\mu_j \sim \mu_0 e^{-L/\xi}$ that decay exponentially with wire length $L$, in practice it is desirable to tune, or at least switch on and off, the couplings in a more controllable fashion.
To this end, consider the TS nanowires in Fig.~\ref{fig:Majoranawires} to be strongly coupled to a ground bulk superconductor, but not fully grounded. For a single wire, both its intrinsic single-electron charging energy $E_c$ and Josephson coupling $E_J$ to the ground bulk SC then become relevant. In the limit of large but finite $E_J/E_c > 1$, one finds an effective parity splitting between even and odd charge states on the wire, directly translating to a Majorana parity splitting for pairs $\chi_L^j,~\chi_R^j$ \cite{vanHeck2012}. On the Hamiltonian level, this term can be incorporated as
\begin{equation}\label{eq:Hinttunable}
H_\mathrm{int, j} = \mu(n_j) i \chi_{L}^j\chi_{R}^j~,~~~
\mu(n_j) = \mu_{0}\cos(\pi n_j)~,
\end{equation}
where $\mu_0$ depends on both $E_c$ and $E_J$ \cite{vanHeck2012}, and $n_j$ is a gate parameter set by a nearby electrostatic gate, thereby controlling the equilibrium charge (parity) on the wire.
We thus see that a nearby collective gate as in Fig.~\ref{fig:Majoranawires}, controlling charge on all nanowires in the device, can simultaneously switch on and off the bilinear Majorana-coupling across all pairs of modes $\chi_{L/R}^j$. Further, at least in principle, one can also address (few) wires individually via additional gates not shown in Fig.~\ref{fig:Majoranawires}.


\section{Measurements in the nanowire device}
\label{AppC}

%

While one can measure some properties of the coupled-wire SYK device with simple tunnel-probes as indicated in Fig.~\ref{fig:Majoranawires}, here we mention another useful capability that comes with the inter-side coupling implementation in Appendix~\ref{AppB}.
For each individually tunable gate voltage in the device, e.g. the collective gate or ones attached to the top/bottom-most nanowires, one can perform projective readouts of the nanowire parities $\hat{q}_j = i\chi_L^j\chi_R^j$ or certain combinations thereof.
The more individually addressable gates, the more completely one may map out the space of parity eigenvalues $q_{j=1,...,N}$. For a detailed discussion of Majorana-parity readout via resonators attached to electrostatic gates, see e.g. Ref.~\cite{Plugge2017,Karzig2017}.\\

To illustrate how this readout works, assume that a gate parameter $n$ in Eq.~\eqref{eq:Hinttunable} is set such that the corresponding interaction is nearly switched off, $n(t) = \frac12 + v(t)$ with a small fluctuating gate voltage $|v(t)| \ll 1$. We then introduce a resonator circuit capacitively coupled to the gate, described by resonator photons $a(t)$. Upon quantizing the fluctuating gate voltage $v(t)$, assuming a capacitive interaction strength $g$ between resonator gate and nanowire, one replaces $v(t) \to [a(t) + a^\dagger(t)]$.
The total nanowire-charge readout setup then is described by
\begin{equation}\label{eq:Hresonator}
H_{\mathrm{readout}} = H_{\mathrm{res}} - g\hat{q}(t)[a + a^\dagger]~,
\end{equation}
where $\hat{q}(t)$ is the time-evolving nanowire charge parity. $H_\mathrm{res}$ here encodes the resonator spectrum, and generates the dynamics for resonator photons $a(t)$.\\
In the strong-coupling regime, leading to a net exchange of resonator
photons with the system, one can directly access e.g. the transmission amplitudes or phase shifts of the resonator that depend on $\langle a(t)\rangle$, and via Eq.~\eqref{eq:Hresonator} also on $\langle \hat{q}(t)\rangle$ \cite{Plugge2017}.
This readout mode hence allows for a direct, time-resolved measurement of Majorana parities $q(t)$ between the two coupled SYK dots in Fig.~\ref{fig:Majoranawires}.

With similar but somewhat more complex measurements, monitoring e.g. the time-dependent resonator photon Green's function $B_{a}(t,t') = \langle a^\dagger(t) a(t')\rangle$ (or any quantity quadratic in photon operators $a,~a^\dagger$), one finds
%
\begin{equation}\label{eq:GFphoton}
B_{a}(t,t') = b_0(t,t') + g^2 \int dt_{1,2} b_0(t,t_1) D_q(t_1,t_2) B_{a}(t_2,t')~~,
\end{equation}
with nanowire-charge correlator $D_q(t,t') = \langle \hat{q}(t)\hat{q}(t')\rangle$. Here $b_0(t,t') = b_0(t-t')$ is a bare photon Green's function of the uncoupled resonator Hamiltonian $H_\mathrm{res}$.
Note that a time-dependent Green's functions as in Eq.~\eqref{eq:GFphoton} is encoded by time- and frequency-resolved resonator occupations $B_{a}(\tau,\omega) = \int d\tau' e^{i\omega\tau'} B_{a}(\tau +\frac{\tau'}{2},\tau -\frac{\tau'}{2})$, that can be measured in principle.
Ignoring back-action of the resonator on the system (that generates an effective interaction as in Eq.~\eqref{eq:Hinttunable}), and to lowest order in charge-resonator interaction $g$, this gives information about $d_0(t,t') = \langle \hat{q}(t)\hat{q}(t')\rangle_0$. The latter expression constitutes a four-Majorana two-sided correlator evaluated with respect to the bare SYK Hamiltonian $H = H^\mathrm{SYK}_L + H^\mathrm{SYK}_R$.

In Sec. IV B of the main text, we discuss how charge or charge-correlator measurements can become useful tools to access OTOCs in coupled-wire SYK dots.

\section{Numerical details}
\label{App:numerics}

\subsection{Exact diagonalization}

The numerical exact diagonalization of the Maldacena-Qi model proceeds in a standard fashion. One first defines a complex fermion basis in order to build the Hilbert space of dimension $2^N$ (for $2N$ Majorana zero-modes). The most convenient choice, as discussed in the main text, is to define complex fermions delocalized accross the two subsystems,
\begin{equation}\label{m4app}
c_j = \frac{1}{\sqrt{2}}(\chi_L^j - i \chi_R^j).
\end{equation}
This basis has two key advantages: first, the anti-unitary time-reversal operator takes the simple form $\Theta = \mathcal{K}$, which makes the Hamiltonian $H_S$ explicitly real and thus saves computational resources; second, the infinite-temperature TFD takes the simple form
\begin{equation}
\ket{\text{TFD}_0} = \ket{0 0 ... 0}.
\end{equation}
reflecting the absence of fermions in all modes $j$.
In practice, one can then generate the TFDs for finite $\beta$ through the relationship \cite{Qi2018}
\begin{equation}
\ket{\text{TFD}_\beta} = \sqrt{\frac{Z_0}{Z_\beta}} e^{-\frac{\beta}{4} H_0} \ket{\text{TFD}_0}
\label{eq:TFD_construction}
\end{equation}
where $H_0 = H_L + H_R$ is the Hamiltonian of the system with zero coupling, $\mu=0$.

This procedure can be made more efficient by implementing the symmetries of the problem: fermion parity $P$ or -- even better -- fermion number modulo 4 ($Q_4$), cf. Ref.~\cite{GarciaGarcia2019}, which includes fermion parity. The Hamiltonian then splits into four blocks of unequal dimensions. The ground state $\ket{G}$ is always found to be in the $Q_4 = 0$ sector, and any thermofield double state constructed using Eq.~(\ref{eq:TFD_construction}) also has $Q_4 = 0$ because $[Q_4, H_0] = 0$.

\subsection{Solution of Schwinger-Dyson equations}

Here we discuss how to solve the large-$N$ Schwinger-Dyson (SD)
equations pertaining to the two coupled SYK models, cf.\ Sec. III-A
Eqs.~\eqref{m6}-\eqref{m8}, in a numerically efficient way. To this
end, we first perform analytical manipulations to implement as many of
their subtle symmetries as possible.

Since we are interested in real-time dynamics of the SD equations, we
work in real frequency $\omega$ and time $t$. To obtain the relevant retarded
propagators we apply the standard analytical
continuation $i\omega_n\to \omega +i\delta$ to Eqs.\ \eqref{m6} and write
\begin{eqnarray}\label{eq:SDeqs}
G^{\rm ret}_{LL}(\omega) &=& \frac{\omega - \Sigma^{\rm ret}_{LL}(\omega)}{D(\omega)}~,\\
G^{\rm ret}_{LR}(\omega) &=& -\frac{i\mu - \Sigma^{\rm
		ret}_{LR}(\omega)}{D(\omega)}~, \nonumber
\end{eqnarray}
with $D(\omega) = \left[\omega-\Sigma^{\rm ret}_{LL}\right]^2 + \left[i\mu-\Sigma^{\rm ret}_{LR}\right]^2$.
The retarded self-energies follow from Eq.~\eqref{m8} as 
\begin{equation}\label{am8}
\Sigma^{\rm ret}_{ab}(t) = J^2G^{\rm ret}_{ab}(t)^3.
\end{equation}
We then employ the method introduced by Banerjee and
Altman~\cite{Altman2016} (supplement S2 therein) to express the self
energies as
\begin{equation}\label{eq:selfE}
\Sigma^{\rm ret}_{ab}(\omega) = -iJ^2\int_0^\infty dt [n_{+-}^2 n_{--} + n_{++}^2 n_{-+}] e^{i\omega t}.
\end{equation}
The factors $n_{ss'}(t)$ are calculated from the spectral
representation of the corresponding propagators $G^{\rm ret}_{ab}(\omega)$ as
\begin{equation}\label{nss}
n_{ss'}(t) = \int_{-\infty}^\infty d\omega \rho_{ab}(s\omega)n_F(s'\omega) e^{-i\omega t}~,
\end{equation}
where $n_F(\omega) = 1/(e^{\beta \omega} +1)$ is the Fermi function.

Given certain symmetries of the spectral functions $\rho_{LL}(\omega)$ and $\rho_{LR}(\omega)$, we now show how to deduce all occupations $n_{ss'}(t)$ from just a single one.
The spectral functions in our case read
\begin{eqnarray}\label{rhos}
\rho_{LL}(\omega) &=& -\frac{1}{\pi}\mathrm{Im}\left[G^{\rm ret}_{LL}(\omega)\right]~,\\
\rho_{LR}(\omega) &=& -\frac{1}{\pi}\mathrm{Im}\left[iG^{\rm
	ret}_{LR}(\omega)\right]~. \nonumber
\end{eqnarray}
While the first line is standard, the $i$ factor on the
second line is unconventional but comes about to give the correct
result for $\rho_{LR}(\omega)$ within our conventions. Consider as an
example the retarded Green's functions for the non-interacting case
\begin{eqnarray}\label{eq:SDansatz}
g^{\rm ret}_{LL}(\omega) &=& \frac{\omega + i\delta}{(\omega + i\delta) - \mu^2}~,\\
g^{\rm ret}_{LR}(\omega) &=& \frac{-i\mu}{(\omega + i\delta) - \mu^2}~, \nonumber
\end{eqnarray}
which solve Eqs.\ \eqref{eq:SDeqs} and \eqref{am8} when $J=0$.
The spectral functions then show Lorentzian peaks (with weight $\frac12$)
\begin{equation}\label{rho_}
\rho_{ab}(\omega) = \frac{1}{2\pi}\left[ \frac{\delta}{(\omega - \mu)^2 + \delta^2} +\sigma_{ab}\frac{\delta}{(\omega + \mu)^2 + \delta^2}\right]~,
\end{equation}
centered at $\omega = \pm \mu$, and are symmetric (anti-symmetric)
around $\omega = 0$, where $\sigma_{LL} = +1$ ($\sigma_{LR} = -1$). 

In our numerical solution we begin from the ansatz
\eqref{eq:SDansatz} with small non-zero $\delta$ and iterate together
with Eqs.\ \eqref{nss} and \eqref{eq:selfE}.
Given that the above symmetry properties persist throughout the iterations of the SD equations, one can relate factors $n_{ss'}(t)$ as
\begin{eqnarray}
n_{--}(t) &=& n_{++}(-t) = [n_{++}(t)]^\ast~,\\
n_{+-}(t) &=& n_{-+}(-t) = [n_{-+}(t)]^\ast~,\notag\\
n_{-+}(t) &=& \sigma_{ab} n_{++}(t)~,\notag\\
n_{--}(t) &=& \sigma_{ab} n_{+-}(t)~\notag.
\end{eqnarray}
The first two of above equations hold since both spectral and Fermi
functions are real, while the last two require the knowledge about the $\omega$-parity $\sigma_{ab}$ of the spectral function. First and third equations yield $n_{--}$ and $n_{-+}$ from $n_{++}$. Inserting the third equation into the second, one further obtains $n_{+-}(t) = \sigma_{ab} [n_{++}(t)]^\ast$.

Hence only $n(t) = n_{++}(t)$ is required to evaluate the retarded
self-energy $\Sigma^{\rm ret}_{ab}(\omega)$. Henceforth we only
consider this quantity and attach to it an index $n_{ab}(t)$ to denote
its relation to the specific spectral function
$\rho_{ab}(\omega)$. Note that $n_{ab}(t)$ is not real in general; it is the
Fourier transform of a real function $\rho_{ab}(\omega)n_F(\omega)$,
so it is Hermitian on the time-axis: $n_{ab}(t) = n_{ab}^\ast(-t)$.

Since we are working with imaginary $G^{\rm ret}_{LR}(\omega)$ (leading also to the unconventional spectral function equation), one has to put back a factor of $-i$ in the self-energy Eq.~\eqref{eq:selfE} above, $\Sigma^{\rm ret}_{LR} \to \tilde{\Sigma}^{\rm ret}_{LR} = -i\Sigma^{\rm ret}_{LR}$. 
With this additional factor, one can also rephrase the self-energies as
\begin{eqnarray}\label{eq:Sigmafinal}
\Sigma^{\rm ret}_{LL}(\omega) &=& -2iJ^2\int_0^\infty dt
\mathrm{Re}\left[n_{LL}^3(t)\right] e^{i\omega t}~,~\\
\tilde{\Sigma}^{\rm ret}_{LR}(\omega) &=& 2iJ^2\int_0^\infty dt
\mathrm{Im}\left[n_{LR}^3(t)\right] e^{i\omega t}~.
\end{eqnarray}
This form makes apparent the explicit anti-Hermiticity of $\Sigma^{\rm ret}_{ab}(\omega)$ on the frequency axis: the argument of the Fourier transform in Eq.~\eqref{eq:Sigmafinal} is real, hence the transformed function is Hermitian. With the prefactors $\mp2iJ^2$, $\Sigma^{\rm ret}_{LL}(\omega)$ and $\tilde{\Sigma}^{\rm ret}_{LR}(\omega)$ are manifestly anti-Hermitian: $\Sigma^{\rm ret}_{LL}(\omega)^\ast = -\Sigma^{\rm ret}_{LL}(-\omega)$ and $\tilde{\Sigma}^{\rm ret}_{LR}(\omega)^\ast = -\tilde{\Sigma}^{\rm ret}_{LR}(-\omega)$.
Since both the bare non-interacting propagator  (terms $\omega$ and $i\mu$
in Eq.~\eqref{eq:SDeqs}) and $\Sigma^{\rm ret}_{ab}(\omega)$ are
anti-Hermitian, the same holds true for any odd-power product of the two
emerging from the Dyson series. Hence also the full propagator  $[G^{\rm
	ret}_{ab}(\omega)]^\ast = -G^{\rm ret}_{ab}(-\omega)$ is
anti-Hermitian on the frequency axis, and $G^{\rm ret}_{ab}(t)$ is
strictly imaginary.

After performing the above manipulations, we further find it convenient to switch to Green's functions and self-energies that diagonalize the SD equation explicitly:
\begin{equation}\label{eq:Gplusminus}
G^{\rm ret}_\pm(\omega) = G^{\rm ret}_{LL} \pm iG^{\rm ret}_{LR}~~,~~~\Sigma^{\rm ret}_\pm(\omega) = \Sigma^{\rm ret}_{LL} \pm i\tilde{\Sigma}^{\rm ret}_{LR}~.
\end{equation}
The spectral densities $\rho_\pm = \rho_{LL}\pm\rho_{LR}$ here follow from the standard relation $\rho_{\pm}(\omega) = -\frac{1}{\pi}\mathrm{Im}G^{\rm ret}_\pm(\omega)$.
From our ansatz in Eq.~\eqref{eq:SDansatz} one sees that $g_+(\omega) = -g_-^\ast(-\omega)$ which implies $\rho_+(\omega) = \rho_-(-\omega)$ initially. One then finds $\Sigma_+(\omega) = -\Sigma_-(-\omega) ^\ast$, and consequently also $G^{\rm ret}_+(\omega) = -G^{\rm ret}_-(-\omega) ^\ast$ which a-priori is not obvious. Hence the symmetry $\rho_+(\omega) = \rho_-(-\omega)$ is kept throughout, and the full solution of the SD equations is encoded in
\begin{equation}\label{Gplus}
[G^{\rm ret}_+(\omega)]^{-1} = \omega - \mu - \Sigma^{\rm ret}_+(\omega),
\end{equation}
where broadening and shift of resonances in $G^{\rm ret}_+(\omega)$ is directly linked to the real/imaginary part of the self-energy
\begin{equation}\label{eq:Sigmaplus}
\Sigma^{\rm ret}_+(\omega) = -2iJ^2\int_0^\infty dt
\left[\mathrm{Re}(n_{LL}^3) - i \mathrm{Im}(n_{LR}^3)\right] e^{i\omega t}~.~
\end{equation}
Here we used
\begin{equation}\label{eq:nLL}
n_{LL/LR}(t) = \int_{-\infty}^{\infty}d\omega \frac{\left[\rho_+(\omega)\pm\rho_+(-\omega)\right]}{2} n_F(\omega) e^{-i\omega t}.
\end{equation}
The physical propagators $G^{\rm ret}_{LL,LR}$ follow from Eq.~\eqref{eq:Gplusminus} and using symmetries, e.g., as $G^{\rm ret}_{LR}(\omega) = -\frac{i}{2}[G^{\rm ret}_+(\omega) + G^{\rm ret}_+(-\omega) ^\ast]$.

We made substantial progress in understanding the structure of the SD
equations for the Maldacena-Qi model and, using symmetries, reduced them to a single
propagator  $G^{\rm ret}_ +(\omega)$ and self-energy
$\Sigma^{\rm ret}_+(\omega)$. Equation \eqref{eq:Sigmaplus} is however highly
nonlinear and Eq.\ \eqref{Gplus} contains an energy scale $\mu$ which makes the solution difficult to access analytically.

To find the fixed-point solution of the SD equations we hence perform
numerical iteration starting from the ansatz for spectral functions
given in Eqs.\ \eqref{rho_} with small nonzero $\delta$.
The self-energy $\Sigma^{\rm ret}_+(\omega)$ is calculated using
Eqs.~\eqref{eq:Sigmaplus}  and \eqref{eq:nLL} with help of fast
Fourier transform algorithms. $G^{\rm ret}_+$ is then computed from
Eq.\ \eqref{Gplus} and used to reconstruct the full retarded
propagator. New spectral functions are then extracted from Eqs.\
\eqref{rhos} and the procedure is restarted from these.
To improve the convergence properties  we follow Ref.\
\cite{Maldacena2016}  and after each round of iteration we mix
the initial propagator with the new solution obtained from the SD equation. 
We declare convergence to a physical solution once the propagators,
spectral densities and, in particular, the energy gap stop changing
within the specified accuracy.
We also check that the solutions are stable with repect to increasing
the number of iterations, frequency/time resolution and cutoffs, and
other non-physical ingredients of the numerical solution, such as the
initial broadening $\delta$.
Results of our numerics are discussed in Sec.\ III of the main text.

\end{document}